\documentclass[11pt]{article}
\pdfoutput=1 
\usepackage{jheppub}
\usepackage{graphicx,verbatim}
\usepackage{psfrag, times}
\usepackage[T1]{fontenc}
\usepackage{epsf}

\def\be{\begin{equation}}
\def\ee{\end{equation}}
\def\bc{\begin{center}}
\def\ec{\end{center}}

\def\bm{{\bf{m}}}

\def\bQ{{\bf{Q}}}

\def\bn{{\bf{n}}}
\def\btau{{\bf{\tau}}}
\def\bl{{\bf{l}}}
\def\bk{{\bf{k}}}

\def\cA{{\mathcal{A}}}

\def\cC{{\mathcal{C}}}
\def\cD{{\mathcal{D}}}

\def\cF{{\mathcal{F}}}
\def\cM{{\mathcal{M}}}

\def\cR{{\mathcal{R}}}

\def\cI{{\mathcal{I}}}

\def\cO{{\mathcal{O}}}
\def\cP{{\mathcal{P}}}
\def\cV{{\mathcal{V}}}
\def\cW{{\mathcal{W}}}

\def\tX{\tilde{X}}
\def\tW{\tilde{W}}

\def\nn{\nonumber}
\def\lam{\lambda}

\def\Del{{\Delta}}

\def\r2{{\sqrt{2}}}

\def\cR{{\mathcal{R}}}

\def\bea{\begin{eqnarray}}
\def\eea{\end{eqnarray}}

\def\bB{{\bf{B}}}

\def\tH{\tilde{H}}
\def\tX{\tilde{X}}

\def\tV{\tilde{V}}

\def\rC{{\rm C}}

\def\hb{\hat{b}}
\def\ttau{\tilde{\tau}}
\def\del{\delta}

\newcommand{\no}{\nonumber}

\begin{document}

\begin{flushright}
KUNS-2710 
\end{flushright}

\title{\Large Towards Spinning Mellin Amplitudes}
\author[]{Heng-Yu Chen${}^{1}$, En-Jui Kuo${}^{1}$ and Hideki Kyono${}^{2}$}
\affiliation{$^1$\rm Department of Physics and Center for Theoretical Sciences, \\
National Taiwan University, Taipei 10617, Taiwan}
\affiliation{$^2$ \rm Department of Physics, Kyoto University,
Kitashirakawa Oiwake-cho, Kyoto 606-8502, Japan}

\emailAdd{heng.yu.chen@phys.ntu.edu.tw}
\emailAdd{ r04222087@ntu.edu.tw }
\emailAdd{ h\_kyono@gauge.scphys.kyoto-u.ac.jp  }
\abstract{We construct the Mellin representation of four point conformal correlation function with external primary operators with arbitrary integer spacetime spins, 
and obtain a natural proposal for spinning Mellin amplitudes.   
By restricting to the exchange of symmetric  traceless primaries, 
we generalize the Mellin transform for scalar case to introduce discrete Mellin variables for incorporating spin degrees of freedom.
Based on the structures about spinning three and four point Witten diagrams, we also obtain a generalization of the Mack polynomial which can be regarded as a natural kinematical polynomial basis for computing spinning Mellin amplitudes using different choices of interaction vertices. 
}

\maketitle

\section{Introduction}\label{Section:Introduction}
\paragraph{}
Recently, Mellin representation of conformal correlation functions, pioneered by Mack \cite{Mack1} 
and further developed for their dual Witten diagrams by Penedones \cite{MellinAmp1} (see also \cite{MellinAmp2}, \cite{PaulosMellin}) has generated significant amount of interests.
In particular, their relatively simple structures have led to dramatic simplification in conformal bootstrap analysis \cite{Gopakumar2016-1, Gopakumar2016-2, Dey-2016},
enable the generalization to superconformal theories \cite{Rastelli-Zhou-1, Rastelli-Zhou-2, Rastelli-Zhou-3, Zhou-1}, 
even improve the loop calculation in Anti-de Sitter space (AdS) \cite{MellinAmp1, LoopMellin, LoopMellin-2, LoopMellin-3, LoopMellin-4, LoopMellin-5}.
It is well-known that scattering amplitudes become ill-defined in conformal field theories due to the lack of asymptotic states, 
however we can still obtain so-called ``Mellin amplitudes'' from the Mellin representation of conformal correlation functions,
and for large-N CFTs possessing holographic duals, these amplitude share many striking common features with the scattering amplitudes in AdS space.
We can view this alternative representation as an integral transform mapping $n$-point conformal correlation functions of $\frac{n(n-3)}{2}$ conformally invariant cross ratios 
into $n$-point Mellin amplitudes of $\frac{n(n-3)}{2}$ Mandelstam-like variables\footnote{
More precisely, here we assume that $n<d$, where $d$ is the number of Euclidean spacetime dimensions, when $n\geq d$, the count of independent cross ratios is changed as $nd - \frac{(d+1)(d+2)}{2}$. 
In this paper we consider four point functions mainly, 
and there are two independent cross ratios or Mellin variables in $d\geq 2$\,.
}.
Mellin amplitudes are meromorphic functions of these Mellin momenta, whose singularities and residues encode respectively the spectrum of CFT operator dimensions and the OPE coefficients.
At these poles, the Mellin amplitudes can factorize into lower point ones, precisely in accordance with the operator product expansion (OPE) of the conformal correlation functions \cite{Factorization}.
\paragraph{}
Moreover, there is a clear separation between the so-called "single" and "double trace" operators\footnote{
Strictly speaking, the terminology of "single" or "double trace" operators only apply to large-$N$ CFTs in the context of AdS/CFT,  
we will refer to operators $\{\cO_{\Del_i}\}$ as the single trace which popularize the spectrum in the generalized free field limit,  while $~\cO_{\Del_1}\partial^J (\partial^2)^m \cO_{\Del_2}$ are referred as double trace operators. More generally, without involving the large-$N$ limit, these may be more appropriately referred as "single twist" or "double twist'' operators.}
in the Mellin representation of conformal correlation functions,
such that the poles within the Mellin amplitudes only correspond to the single trace operators. While the singularities associated with the double trace operators are encoded in a separate product of $\Gamma$-functions which can be regarded as part of the transformation Mellin kernel.
This separation of operator spectrum resembles large-$N$ expansion and is particularly useful for those CFTs with weakly coupled Anti-de Sitter (AdS) space dual.
In this note, we will focus mostly on the four point conformal correlation functions involving both external scalar and spinning primary operators,
their kinematical parts which can be fixed by conformal symmetries, are known as conformal block/partial waves.
Conformal partial waves are mapped into the Mellin partial amplitudes via Mellin transformation, they account for most of the complexities in the full amplitudes.
While the remaining dynamical information, such as the spectrum of operator dimensions and OPE coefficients, 
can be obtained from the corresponding Witten diagram computations in AdS space, we often package these into so-called spectral function.
\paragraph{}
The four point scalar Mellin amplitude, as will be reviewed shortly in Section \ref{Sec:Scalar Mellin}, can be derived from a single four point Witten exchange diagram computation for each exchange channel,
to derive the Mellin amplitude involving external operators with spins however, few complications arise.
First there can be exchange primary operators transforming under different irreducible representations of the spacetime rotation group, 
this is in contrast with the scalar case where we can only have the symmetric traceless exchange.
Moreover for each irreducible representation, there can be multiple possible interaction vertices in AdS space,
this is consistent with the fact that spinning conformal correlation functions can be expanded into multiple independent spacetime tensor structures determined by conformal symmetries.
\paragraph{}
In this note, we aim to make a modest step towards constructing the full spinning Mellin amplitudes, 
our approach here is first restricting ourselves to only the exchange of symmetric traceless primary operators,
admittedly this restriction only yields partial contributions in general dimensions, however this is actually sufficient to give the complete contributions in three dimensions\footnote{Recently there is an interesting work considering Mellin amplitude with external fermionic operators \cite{FermionicMellin}.}.
The corresponding AdS space three point interaction vertices for their dual tensor fields  
have been classified systematically in \cite{3pt-Coupling}, \cite{SpinningWitten}.
The resultant spinning three point Witten diagrams, which are building blocks of the four point exchange diagrams, 
can be expanded in terms of kinematical CFT tensor basis, their dynamical expansion coefficients contain the singularities encoding the spectrum of double trace operators. 
We review the relevant details in Section \ref{Sec:SpinningThreePoint}.
Since the spinning Mellin amplitude should only contain the singularities associated with the single trace operators, 
we will focus on the kinematical contributions, they are independent from the aforementioned vertex-basis dependent dynamical factors which only affect the double trace operator spectrum. 
\paragraph{}
Our main results are summarized in \eqref{Spinning4pt-Mellin}, \eqref{SpinningSpecFunc} and \eqref{SpinningMellinAmp}, 
they can be regarded as the Mellin representation of an interaction vertex choice-dependent contribution to the four point spinning correlation function.
In deriving these results, we extended the Symanzik star-formula {using the formula in Appendix of \cite{generalized Symanzik}} and generalized the Mellin transform to include discrete Mellin variables in order to incorporate the discrete spacetime spin degrees of freedom,
and we proposed a natural generalization of Mack polynomial to express the resultant spinning Mellin amplitudes.
While for general dimensionality, these results represent only the partial contributions to the full spinning Mellin amplitudes. However for three dimensions, 
only symmetric traceless exchanges are allowed, once the basis for the interaction vertices is chosen, our results can be used to constructed the full spinning Mellin amplitudes.
\paragraph{}
We relegate most of the computational details, some useful formulae and identities in few appendices.

\section{Scalar Mellin Amplitudes from Witten Diagrams}\label{Sec:Scalar Mellin}
\paragraph{}
In this section, we will slightly generalize the derivation for the Mellin representation of four point correlation of scalar primary operators,
using the scalar Witten diagram approach given in \cite{MellinAmp1, MellinAmp2, PaulosMellin}, closely related computations were also given in \cite{Mack1} and \cite{DO-2011}.
\paragraph{}
Let us first parameterize the four point scalar correlation function in (Euclidean) $d$-dimensional conformal field theory (CFT) into conformal partial wave expansion following \cite{ConformalRT}\footnote{In this paper, unless otherwise stated, we will be working in embedding formalism such that $\{P_i^A\} \in {\mathbb M}^{1, d+1}$, $P_i^2 =0$ are embedding space coordinates etc.\,. See for example \cite{Rychkov-Lecture} for a good introduction to embedding formalism in CFT. }:
\bea\label{4ptScalarFunction}
\langle\cO_{\Del_1}(P_1)\cO_{\Del_2}(P_2)\cO_{\Del_3}(P_3)\cO_{\Del_4}(P_4) \rangle
&=&\cP(P_i)\sum_{J=0}^{\infty} \int^{+\infty}_{-\infty} d\nu\, b_J(\nu) \cF_{\nu, J} (u, v)\nn\\
&=& \sum_{J=0}^{\infty} \int^{+\infty}_{-\infty} d\nu\, b_{J}(\nu)\cW_{\nu, J} (P_i),
\eea
\be
\cP(P_i)=\frac{1}{(P_{12})^{\frac{\Del_1+\Del_2}{2}}(P_{34})^{\frac{\Del_3+\Del_4}{2}}} {\left(\frac{P_{24}}{P_{14}}\right)^{\frac{\Del_{12}}{2}} \left(\frac{P_{14}}{P_{13}}\right)^{\frac{\Del_{34}}{2}}},
\ee
where $\cO_{\Del_i}(P_i), i=1,2,3,4$ are scalar primary operators of scaling dimensions $\Del_i$,
$P_{ij} = -2P_i\cdot P_j$, $\Del_{ij} =\Del_i-\Del_j$, $i, j =1, \dots, 4$ and $(u, v)$ are conformally invariant cross ratios:
\be\label{CrossRatios}
u = \frac{P_{12} P_{34}}{P_{13} P_{24}}, \quad v = \frac{P_{14}P_{23}}{P_{13}P_{24}}.
\ee
In \eqref{4ptScalarFunction}, we have chosen the $\cO_{\Del_{1,2}}/\cO_{\Del_{3,4}}$ OPE channel, 
and we have used the so-called {\it ``spectral representation''}, where $\nu$ is the so-called spectral parameter.
This representation splits the correlation function into purely kinematical part fixed by the conformal symmetries, as encoded in conformal block $\cF_{\nu, J}(u, v)$ which is an eigenfunction of scalar quadratic Casimir equation \cite{DO-2003, DO-2011}, while the remaining $b_J(\nu)$ is the so-called spectral function containing the dynamical information of a given conformal field theory.
\paragraph{}
It is known that $\cF_{\nu, J}(u, v)$ and ``conformal partial wave'' $\cW_{\nu, J}(P_i)$\footnote{Please see discussion below regarding the subtlety in the choice of parameterization.} also admit the following Mellin representations:
\bea\label{CB-Mellin}
\cF_{\nu, J}(u, v) &=&  \int^{+i\infty}_{-i\infty} \frac{dt ds}{(4\pi i)^2}\, \cM_{\nu, J}(s, t) u^{\frac{t}{2}} v^{-\frac{(s+t)}{2}} \prod_{i<j}\Gamma(\delta_{ij}),\\
\label{CPW-Mellin}
\cW_{\nu, J}(P_i) &=& \int^{+i\infty}_{-i\infty} \frac{dt ds}{(4\pi i)^2} \cM_{\nu, J} (s, t)\prod_{i< j} {\Gamma(\delta_{ij})}{P_{ij}^{-\delta_{ij}}},
\eea
where $(s, t)$ are the so-called ``Mellin momenta'' and $\cM_{\nu, J}(s, t)$ is the kinematical Mellin partial amplitude, which can be regarded as the Mellin transformation of the conformal partial wave.
Here the parameters $\delta_{ij} = \delta_{ji}$, $i\neq j$ appearing in the kernel of Mellin transformation are explicitly given as:
\bea\label{Mellin-deltaij}
&&\delta_{12} = \frac{\Delta_1+\Delta_2-t}{2}, \quad  \delta_{34} = \frac{\Delta_3+\Delta_4-t}{2},\quad \delta_{13}=\frac{\Delta_{34}-s}{2},\quad \delta_{24} =- \frac{\Delta_{12}+s}{2},\nn\\
&&\delta_{14} = \frac{s+t+\Delta_{12}-\Delta_{34}}{2}, \quad \delta_{23} = \frac{s+t}{2},
\eea
and they satisfy the constraints
\footnote{
Here we have adopted the parameterization of Mellin variables $(s, t)$ following the conventions in \cite{ConformalRT}, 
however we can re-parameterize them to turn into manifestly crossing invariant form: $s \to -s'-t'+\Delta_2+\Delta_3$ and $t \to s'$, 
such that the factor $\prod_{i<j}\Gamma(\delta_{ij})$ becomes crossing symmetric \cite{Rastelli-Zhou-2}: 
\bea
\prod_{i<j}\Gamma(\delta_{ij})&=&
\Gamma\left(\frac{\Delta_1+\Delta_2-s'}{2}\right)\Gamma\left(\frac{\Delta_3+\Delta_4-s'}{2}\right)
\Gamma\left(\frac{\Delta_1+\Delta_4-t'}{2}\right)\\
&&\qquad \qquad \times \Gamma\left(\frac{\Delta_2+\Delta_3-t'}{2}\right)
\Gamma\left(\frac{\Delta_1+\Delta_3-u'}{2}\right)\Gamma\left(\frac{\Delta_2+\Delta_4-u'}{2}\right)\nn\,,
\eea 
where we introduced a new Mellin variable $u'$ through the relation: $s'+t'+u' = \sum_{i=1}^4 \Del_i$. 
}:
\be\label{deltaij-constraint}
\sum_{j (\neq i)} \delta_{ij} = \Del_i, \quad i =1, 2, 3, 4.
\ee
Hence for a given set of scaling dimensions $\{\Del_i\}$, there are in fact only two independent parameters in four point Mellin amplitude,
denoted as $(s, t)$\footnote{ 
We can solve these constraints \eqref{deltaij-constraint} by introducing the ``fictitious momenta'' $\{p_i\}$ satisfying $\sum_{i=1}^4 p_i = 0$ and $p_i^2 = -\Del_i$,
then $\delta_{ij} = p_i\cdot p_j$ automatically satisfy \eqref{deltaij-constraint}. 
We can regard $\{p_i\}$ as the momenta of the external particles involved in the four point Mellin amplitude \eqref{MellinAmp}, with rest mass $p_i^2 = -\Del_i$, 
the Mellin momenta $t = -(p_1+p_2)^2$ and $s = -(p_1+p_3)^2$ are in fact Mandelstam-like variables in this parameterization.}.
The integrand $\cM_{\nu, J}(s, t) $ in \eqref{CB-Mellin} and \eqref{CPW-Mellin} can be further decomposed into two parts:
\bea\label{MellinPartial}
\cM_{\nu, J}(s, t) &=&  {\omega}_{\nu, J}(t) P_{\nu, J}(s, t),
\eea
where the purely $t$-dependent part is given by:
\be\label{Omega-t}
{\omega}_{\nu, J}(t) =\frac{\Gamma\left(\frac{\Del_1+\Del_2+J-h\pm i\nu}{2}\right) \Gamma\left(\frac{\Del_3+\Del_4+J-h\pm i\nu}{2}\right)}{8\pi \Gamma(\pm i\nu)}
\frac{\Gamma\left(\frac{h-J-t\pm i\nu}{2}\right) }{\Gamma\left(\frac{\Del_1+\Del_2-t}{2}\right) \Gamma\left(\frac{\Del_3+\Del_4-t}{2}\right)},
\ee
{where $h = \frac{d}{2}$, and we introduced a short-hand notation: $\Gamma(a\pm b)\equiv \Gamma(a+b)\Gamma(a-b)$.}
The remaining factor $P_{+\nu, J}(s, t) = P_{-\nu, J}(s, t)$ is the so-called Mack polynomial of degree $J$ in both $t$ and $s$ \cite{Mack1},
which is a smooth polynomial and its form will be derived momentarily.
\paragraph{}
The full Mellin amplitude $\cM(s, t)$ is obtained upon integrating $\cM_{\nu, J}(s, t)$ over the spin-dependent dynamical spectral function $b_{J}(\nu)$:
\be\label{MellinAmp}
\cM(s, t) = \sum_{J=0}^{\infty} \int^{+\infty}_{-\infty} d\nu\, b_J(\nu)  \cM_{\nu, J}(s, t),
\ee 
and we can succinctly write down the Mellin representation of four point scalar correlation function:
\be\label{MellinTrans-4ptScalar}
\langle\cO_{\Del_1}(P_1)\cO_{\Del_2}(P_2)\cO_{\Del_3}(P_3)\cO_{\Del_4}(P_4) \rangle 
=\int^{+i\infty}_{-i\infty}\frac{ds dt}{(4\pi i)^2} \cM(s, t) \prod_{i< j} \Gamma(\delta_{ij}) P_{ij}^{-\delta_{ij}}.
\ee
As discussed in details in \cite{ConformalRT}, 
$\cM(s, t)$ has very specific singularity structures. 
First, it does not contain any poles in $s$ or $t$  corresponding to the exchange of the so-called double-trace operators of the generic form  
$\sim \cO_{\Del_1}\partial^J (\partial^2)^m \cO_{\Del_2}$ in the sense of large-$N$ counting, 
with the parameterization \eqref{MellinTrans-4ptScalar},
the relevant singularities are contained in the explicit $\Gamma$-functions in the remaining kernel.
Instead $\cM(s, t)$ only contains the singularities corresponding to the exchanges of the single-trace operators in the OPE channels, 
which are necessary to reproduce the correct small cross ratio expansion of the conformal block in \eqref{CB-Mellin}.
More explicitly, for a common exchange primary operator $\cO_{\Del, J}$ of twist $\Del-J$ in the $\cO_{\Del_{1,2}}/\cO_{\Del_{3,4}}$ OPE channels, $\cM(s, t)$ has infinite number of simple poles at $t = \Del - J + 2m$, $m = 0, 1, 2, 3, \dots$, 
and the following Laurent expansion: 
\be\label{MellinExpansion}
\cM(s, t) =  C_{12\cO_{\Del, J}}C_{34\cO_{\Del, J}}\sum_{m=0}^{\infty}\frac{ \bQ_{J, m}(s)}{t-(\Del-J+2m)} + \cR_{J-1} (s, t),
\ee
where the purely kinematical residue function is given by\footnote{
Here we followed the pole structure of the spectral function $b_J(\nu)$ given in \cite{ConformalRT}:
\be
b_J(\nu)\sim C_{12\cO_{\Del, J}}C_{34\cO_{\Del, J}}\frac{K_{\Delta,J}}{\nu^2 + (h-\Delta)^2}\,,
\ee
where
\be
K_{\Delta,J}\equiv\frac{
\Gamma(\Delta+J)
\Gamma(\Delta-h+1)
(\Delta-1)_J
}
{
4^{J-1}
\Gamma \left(\frac{\Delta+J\pm \Delta_{12}}{2}\right)
\Gamma \left(\frac{\Delta+J\pm \Delta_{34}}{2}\right)
\Gamma \left(\frac{\Delta_1+\Delta_2+J-h\pm (h-\Delta)}{2}\right)
\Gamma \left(\frac{\Delta_3+\Delta_4+J-h\pm (h-\Delta)}{2}\right)
}\,.
\ee
}
\bea\label{Def:boldQ-function}
&&\bQ_{J, m}(s) =\\
&&-\frac{2\Gamma(\Del+J)(\Del-1)_J}{4^J\Gamma\left(\frac{\Del+J\pm\Del_{12}}{2}\right) \Gamma\left(\frac{\Del+J\pm\Del_{34}}{2}\right)}
\frac{Q_{J, m}(s)}{m!(\Del-h+1)_m \Gamma\left(\frac{\Del_1+\Del_2-\Del+J-2m}{2}\right) \Gamma\left(\frac{\Del_3+\Del_4-\Del+J-2m}{2}\right)},\nn\\[10pt]
&& Q_{J, m}(s) = P_{-i(\Del-h), J}(s, t = \Del-J+2m).\label{Def:Q-function}
\eea
Here $C_{12\cO_{\Del, J}}$ and $C_{34\cO_{\Del, J}}$ are the OPE coefficients for $\cO_{\Del, J}$, and its descendants labeled by
the infinite number of poles $m>0$. $Q_{J, m}(s)$ is a polynomial of degree $J$ in $s$, whose explicit form can be obtained directly from Mack polynomial.
The remaining regular piece $\cR_{J-1}(s, t)$ is a degree $J-1$ polynomial in $(s, t)$.
{Its origin is basically the Laurent expansions of the gamma functions, 
and the contact interactions generated by different choices of interaction vertices also can contribute to this term
when considering the corresponding Witten diagrams to be reviewed next.}
\paragraph{}
In the context of the AdS/CFT correspondence, it is well known that four point scalar correlation function \eqref{4ptScalarFunction} can be reproduced from summing over the contributions from the four point scalar contact and exchange Witten diagrams (WD) in Anti-de Sitter space (AdS) \cite{SpinningAdS, SpinningAdS-1, SpinningAdS-2}. 
Moreover, it was known recently that we can further decompose these Witten diagrams into kinematical building blocks known as ``geodesic Witten diagrams'' (GWD) \cite{ScalarGWD}, 
which was shown to reproduce conformal partial wave up to overall factors.
In particular, the contact Witten diagrams can be decomposed into infinite tower of geodesic Witten diagrams corresponding to the exchange of double trace operators, 
while the exchange Witten diagrams for a given channel, which will be discussed extensively next, consists of geodesic Witten diagrams for the single and double trace operators. 
The clean separation of single and double trace operator exchanges in Witten diagrams clearly mimics the Mellin representation of the CFT scalar correlation functions,
and this strongly implies we can derive scalar Mellin amplitude $\cM(s, t)$ holographically from Witten diagram/geodesic Witten diagram computations.
\paragraph{}
To clarify, in \eqref{4ptScalarFunction} we have followed the parameterization of the scalar correlation function in \cite{ConformalRT}, where  $\cW_{\nu, J}(P_i)$ can contain $\Gamma$-function pre-factors with singularities corresponding to the double trace operators. This differs from the usual definition of conformal partial wave, they are therefore more appropriately computed through exchange Witten diagrams. When we rewrite $\cW_{\nu, J}(P_i)$ into Mellin representation however, these singularities are repackaged into the definition of the Mellin transformation kernel \eqref{CPW-Mellin} such that $\cM_{\nu, J}(s, t)$ does not contain double trace contributions. 
In this sense, we can regard $\cM_{\nu, J}(s,t)$ as the natural  Mellin space basis for expanding scalar Mellin amplitude.
The basic strategy here is therefore that, we will use the cutting identity of the bulk to bulk propagator \eqref{Cutting Identity} to cut the four point exchange Witten diagram into two copies of three point Witten diagrams, 
whose form itself can be fixed by AdS isometries to be proportional to the CFT three point function up to an overall dynamical factor, which can be attributed to the spectral function $b_J(\nu)$.
As we explained in the previous paragraph, the Witten diagrams can be decomposed into geodesic Witten diagrams, we shall comment on how Mellin amplitude can also be reproduced using geodesic Witten diagrams at the end of section.
\paragraph{}
Later, we will see that the expression \eqref{MellinPartial} can be reproduced 
by a single four point exchange Witten diagram with a bulk AdS tensor field dual to the primary operator $\cO_{\Del, J}$
\footnote{These individual Witten diagrams were called  
Witten blocks in \cite{Gopakumar2016-1,Gopakumar2016-2}.}.
Comparing with the conventional Witten diagram computations where we need to sum over all possible exchange channels to reproduce the full correlation functions,
and the final expression is manifestly crossing invariant \cite{DHokerFreedman},
here we consider only the Witten diagram for $(12)(34)$ exchange channel.
The justification for this is that as argued in \cite{ConformalRT}, 
in this case only this channel contains the contribution from the single trace operators (plus other double trace operators) as needed for Mellin amplitude,
other $(13)(24)$ and $(14)(23)$ channels contain only double trace operators contributing to the $\Gamma$-functions in the transformation kernel.

\begin{figure}[t]\centering
\includegraphics[width=7.5cm]{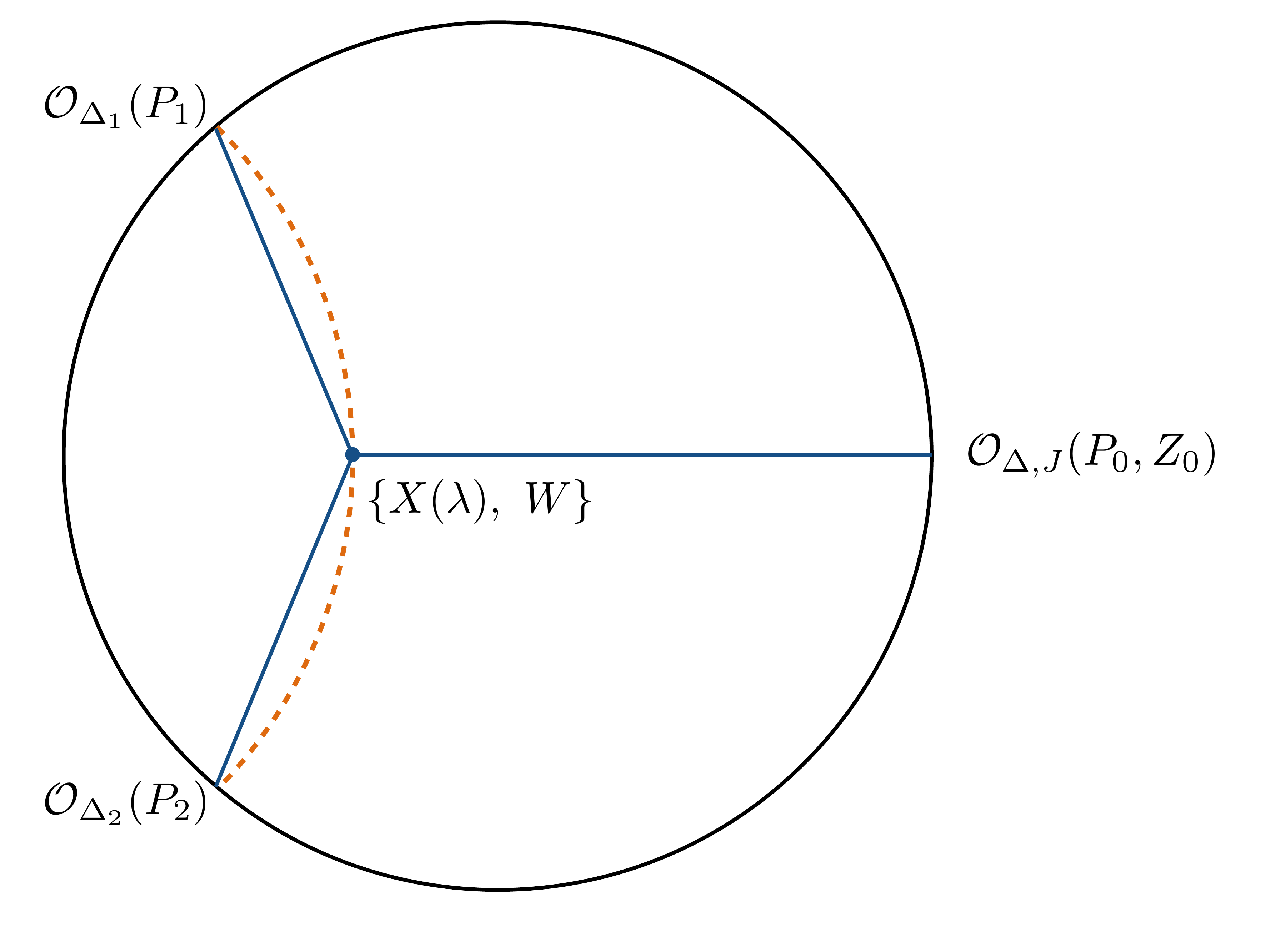}
\caption{{Three point geodesic Witten diagram where the orange curve describes the geodesic $\Gamma_{12}$, 
and the interaction vertex moves only on the geodesics.
}  \label{Fig:3ptGWD}
  }
\end{figure}

More generally, in deducing the Mellin amplitude, 
we have to sum over contributions coming from the other different exchange channels if they also contain contributions of single trace operators \cite{AldayMellin}, 
our computation here can be easily recycled.
\paragraph{}
Let us start with the construction of three point Witten diagram, we consider the holographic dual interaction among a pair of AdS scalars and a spin-$J$ divergent free tensor field,
they interact through following vertex \cite{SpinningAdS, SpinningAdS-1}:
\be\label{SST-3ptVertex}
g_{\Phi_1 \Phi_2 \Xi}\int_{\rm AdS_{d+1}} dX\,  \Phi_1 (X)\nabla_{A_1}\dots \nabla_{A_J} \Phi_2(X) \Xi^{A_1 \dots A_J}(X) 
\ee
which is unique up to the equation of motion. Here $g_{\Phi_1 \Phi_2 \Xi}$ is the coupling constant.
This yields the following three point Witten diagram calculation:
\bea
&&\langle\cO_{\Del_1}(P_1)\cO_{\Del_2}(P_2)\cO_{\Del, J}(P_0, Z_0)\rangle_{\rm WD}\nn\\ 
&&=\frac{g_{\Phi_1\Phi_2\Xi}}{\sqrt{\cC_{\Del_1, 0}\cC_{\Del_2,0}\cC_{\Del, J}}} \int_{\rm AdS_{d+1}} dX\, \Pi_{\Del_1, 0}(X, P_1)\frac{ \Pi_{\Del, J}(X, P_0; K, Z_0) (W\cdot \nabla)^J \Pi_{\Del_2, 0}(X, P_2)}{J!\left(h-\frac{1}{2}\right)_J}\nn\\
&& = \frac{g_{\Phi_1\Phi_2\Xi}}{\sqrt{\cC_{\Del_1, 0}\cC_{\Del_2,0}\cC_{\Del, J}}} b(\Del_1, \Del_2, \Del, J)
 \begin{bmatrix}
				\Delta _1 &\Delta _2&  \Del    \\[0.05em]
				0 & 0           & J \\[0.05em]
				0         &     0     & 0  
			\end{bmatrix},	
\label{Def:SST-3ptWD}
\eea
where $\cO_{\Del, J}(P_0, Z_0)$ is a spin-$J$ symmetric traceless tensor primary operator 
of scaling dimension $\Del$ 
contracted with the polarization vector $Z_0$.
The differential operator $K$ is introduced to contract with the bulk polarization vector $W$\footnote{
For the details such as the explicit form of $K$, for example, please see \cite{SpinningAdS} or \cite{SpinningGWD}.}.
The square parentheses is a three point function including tensor structure defined in Sec.\ref{Sec:Spinning Mellin}.
Here we have introduced the AdS bulk to boundary propagator for spin-$J$ bulk tensor field:
\be\label{Def:Spin l-bulk-boundary}
\Pi_{\Del, J}(X, P; W, Z) = \cC_{\Del, J}\frac{[2(X\cdot Z)(P\cdot W)-2(X\cdot P)(Z \cdot W)]^{J}}{(-2P \cdot X)^{\Del+J}} 
\ee
and normalization constant is
\be\label{Def:CDel-J}
\cC_{\Del, J} = \frac{(J+\Del-1)\Gamma(\Del-1)}{2\pi^{h}\Gamma(\Del+1-h)}.
\ee
The final result in \eqref{Def:SST-3ptWD} is proportional to the unique scalar-scalar-tensor CFT three point function given in the box basis, whose explicit form is given in \eqref{CFT-BoxBasis}.
The explicit integration of interacting vertex over the entire $d+1$-dimensional AdS space yields the dynamical factor:
\be\label{Def:bJnu}
b(\Del_1, \Del_2, \Del, J) = 2^{J-1}(-1)^J \pi^h {\cC_{\Del_1,0}\cC_{\Del_2, 0}\cC_{\Del, J}} \frac{\Gamma\left(\frac{\Del_1+\Del_2+J-h\pm (h-\Del)}{2}\right)\Gamma\left(\frac{\Del+J\pm \Del_{12}}{2}\right)}{\Gamma(\Del_1)\Gamma(\Del_2)\Gamma(\Del+J)},
\ee
and we can identify $\frac{g_{\Phi_1\Phi_2\Xi}}{\sqrt{\cC_{\Del_1, 0}\cC_{\Del_2,0}\cC_{\Del, J}}} b(\Del_1, \Del_2, \Del, J)$ as the OPE coefficient $C_{12\cO_{\Del, J}}$.
When we later promote $\Del$ into continuous spectral parameter, 
the pre-factor $b(\Del_1, \Del_2, \Del. J)$ has singularities at special values $\Del=\Del_1+\Del_2+J+2m$, $m=0, 1, 2, 3,\dots$ corresponding to the double trace operators or their descendants, these are responsible for generating the double trace operator contributions to the four point exchange Witten diagram.
\paragraph{}
To separate the dynamical and kinematical parts in three point function computation, in \cite{SpinningGWD} the authors introduced the three point ``geodesic Witten diagram'' (GWD) where the interaction vertex is now restricted to move instead along the AdS-geodesic $\Gamma_{12}$ connecting two out of the three boundary points where the CFT operators are inserted, in this case $P_1$ and $P_2$ (see Fig.\ref{Fig:3ptGWD}). 
This yields the following integral:
\bea
 &&\langle\cO_{\Del_1}(P_1)\cO_{\Del_2}(P_2)\cO_{\Del, J}(P_0, Z_0)\rangle_{\rm GWD}\nn\\ 
&&=\frac{g_{\Phi_1\Phi_2\Xi}}{\sqrt{\cC_{\Del_1, 0}\cC_{\Del_2,0}\cC_{\Del, J}}} \int_{\Gamma_{12}} d\lambda \, \Pi_{\Del_1, 0}(X(\lambda), P_1)\frac{ \Pi_{\Del, J}(X(\lambda), P_0; K, Z_0) (W\cdot \nabla)^J \Pi_{\Del_2, 0}(X(\lambda), P_2)}{J!\left(h-\frac{1}{2}\right)_J}\nn\\
&& = \frac{g_{\Phi_1\Phi_2\Xi}}{\sqrt{\cC_{\Del_1, 0}\cC_{\Del_2,0}\cC_{\Del, J}}} \hb(\Del_1, \Del_2, \Del, J)
 \begin{bmatrix}
				\Delta _1 &\Delta _2&  \Del    \\[0.05em]
				0 & 0           & J \\[0.05em]
				0         &     0     & 0  
			\end{bmatrix},	
\label{Def:SST-3ptGWD} 
\eea
where $\lambda$ is the line parameter of the geodesic $\Gamma_{12}$ and the overall dynamical factor now becomes:
\bea
\hb(\Del_1, \Del_2, \Del, J) &=& 2^{J-1}(-1)^J (\Del_2)_J \cC_{\Del_1, 0}\cC_{\Del_2, 0}\cC_{\Del, J}  \frac{\Gamma\left(\frac{\Del+J\pm \Del_{12}}{2}\right)}{\Gamma(\Del+J)}\nn\\
&=& 2^{J-1}(-1)^J (\Del_2)_J \cC_{\Del_1, 0}\cC_{\Del_2, 0} \frac{\Gamma\left(\frac{\Del+J\pm\Del_{12}}{2}\right)}{2\pi^h \Gamma(\Del-h+1)(\Del-1)_J} \label{Def:hbJnu}.
\eea
Here we can see that \eqref{Def:hbJnu} is merely a non-divergent overall factor multiplying the CFT box basis.
\paragraph{}
Next we consider constructing the full four point Witten diagram using the three point Witten diagrams we just reviewed, 
this also enables us to derive both partial and full Mellin amplitudes \eqref{MellinPartial} and \eqref{MellinAmp}.
The starting point is the cutting identity of bulk-bulk spin $J$ propagator:
\bea\label{Cutting Identity}
&&\Pi_{\Del, J}(X, \tX; W, \tW) = \\
&&\quad \int^{+\infty}_{-\infty} d\nu \int_{\partial{\rm AdS_{d+1}}} dP_0\frac{\nu^2}{\nu^2+(\Del-h)^2} \frac{ \Pi_{h+i\nu, J}(X, P_0; W, \cD_{Z_0}) \Pi_{h-i\nu, J}(\tX, P_0; \tilde{W}, Z_0)}{\pi J! (h-1)_J},\nn
\eea
which is valid for general bulk points $(X, \tilde{X})$.
{This identity means that a bulk-bulk propagator can be cut into two bulk-boundary propagators, 
and in the first bulk-boundary propagator, the polarization vector $Z_0$ is replaced with a differential operator $\cD_{Z_0}$ defined in \eqref{Def:DZ0} to take a contraction.}
We can evaluate the four point exchange Witten diagram $W^{\text{4pt}}_{\Del, J} (P_i)$ for the bulk dual field of $\cO_{\Del, J}$ using the previous three point results (see Fig.\ref{Fig:4ptSplit}): 
\begin{figure}[h]\centering
\includegraphics[width=15cm]{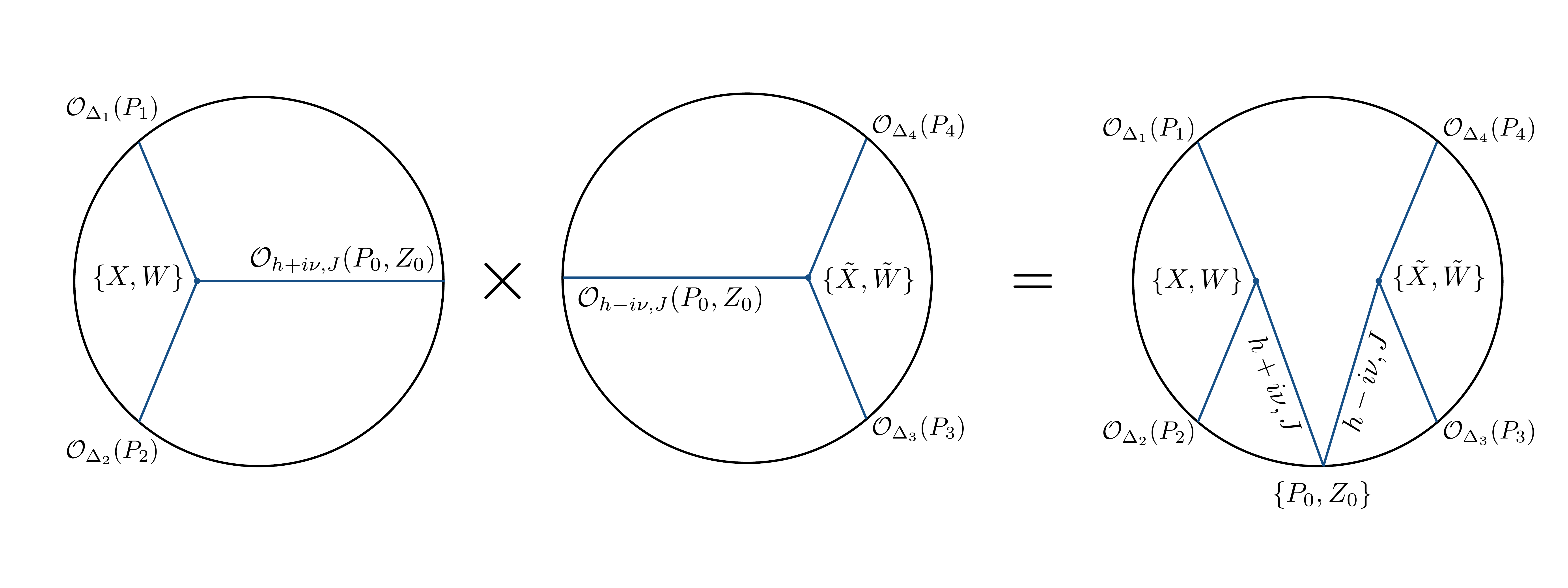}
\caption{{A four point exchange diagram can be constructed by gluing two three point diagrams through the cutting identity of bulk-bulk propagator.
}  \label{Fig:4ptSplit}
  }
\end{figure}
{
\bea
W^{\text{4pt}}_{\Del, J} (P_i) &=& \frac{g_{\Phi_1\Phi_2\Xi} g_{\Phi_3\Phi_4\Xi}}{\left[J!(h-\frac{1}{2})_J\right]^2}\int_{\rm AdS_{d+1}} dX d\tilde{X}\,
\Pi_{\Del_1, 0}(X, P_1){(K\cdot \nabla)^J \Pi_{\Del_2, 0}(X, P_2)}\\
&& 
\qquad\qquad \times 
\Pi_{\Del, J}(X, \tX; W, \tW)\,  \Pi_{\Del_3, 0}(\tilde{X}, P_3){(\tilde{K}\cdot \tilde{\nabla})^J \Pi_{\Del_4, 0}(\tilde{X}, P_4)}\nn\\
&=& \frac{1}{J!(h-1)_J}\int^{+\infty}_{-\infty} \frac{d\nu}{2\pi} \frac{2 \nu^2}{\nu^2+(\Del-h)^2}
\left[\cC_{h\pm i\nu,J} \prod_{i=1}^4\cC_{\Delta_i,0}\right]^{\frac{1}{2}}
\int_{\partial{\rm AdS_{d+1}}} dP_0\nn\\
&&\qquad \times 
\langle\cO_{\Del_1}(P_1)\cO_{\Del_2}(P_2)\cO_{h+i\nu, J}(P_0, \cD_{Z_0})\rangle_{\rm WD}
\langle\cO_{\Del_3}(P_3)\cO_{\Del_4}(P_4)\cO_{h-i\nu, J}(P_0,Z_0)\rangle_{\rm WD}\,.\nn
\eea
}
Putting various pieces together, we can parameterize the four point scalar Witten diagram as:
\be\label{4pt-Witten2}
W^{\text{4pt}}_{\Del, J} (P_i) = {g_{\Phi_1\Phi_2\Xi}g_{\Phi_3\Phi_4\Xi}}
\int^{+\infty}_{-\infty} \frac{d\nu}{2\pi}  \frac{2 \nu^2}{\nu^2+(\Del-h)^2} b(\Del_1, \Del_2,h+i\nu , J) b(\Del_3, \Del_4, h-i\nu, J)~ \cI_{\nu,J}(P_i)\,,
\ee
where we have separated the dynamical information as encoded in the first three terms of the integrand above, 
as they contain singularities in the complex $\nu$ plane and partly will become spectral function $b_J(\nu)$. 
While the the purely kinematic part is encoded in the following kinematic integral $ \cI_{\nu,J}(P_i)$\footnote{Here the differential operator $\cD_{Z_0^A}$ is defined as \cite{Thomas-D}:
\be\label{Def:DZ0}
\cD_{Z_0^A} = \left(h-1+Z_0\cdot \frac{\partial}{\partial Z_0}\right)\frac{\partial}{\partial Z_0^A} - \frac{1}{2} Z_{0 A}\frac{\partial^2}{\partial Z_0 \cdot \partial Z_0}.
\ee
 } : 
\bea\label{Def:ScalarKI}
 \cI_{\nu,J}(P_i)&\equiv& \frac{1}{J! (h-1)_J}\int_{\partial{\rm AdS_{d+1}}} dP_0
\begin{bmatrix}
				\Delta _1 &\Delta _2&  h+i\nu    \\[0.05em]
				0 & 0           & J \\[0.05em]
				0         &     0     & 0  
\end{bmatrix}
\cdot 
\begin{bmatrix}
				h-i\nu &\Delta _3& \Del_4      \\[0.05em]
				J & 0           & 0 \\[0.05em]
				0         &     0     & 0  
\end{bmatrix}\\
&=&
\frac{1}{P_{12}^{\gamma_{12}}P_{34}^{\gamma_{34}}}
\int_{\partial{\rm AdS_{d+1}}} dP_0
\left(\prod_{i=1}^4 \frac{1}{P_{0i}^{\gamma_{0i}}}\right)
 \frac{1}{J! (h-1)_J}
 (2 P_0\cdot \cP_{12} \cdot \cD_{Z_0})^J (2 P_0\cdot \cP_{34}\cdot Z_0)^J\,,\nn
\eea
where $\cP_{12}$ and $\cP_{34}$ are defined as:
\bea
\cP_{12}^{AB}\equiv P_{1}^AP_2^B-P_{2}^AP_1^B\,,\quad \cP_{34}^{AB}\equiv P_{3}^AP_4^B-P_{4}^AP_3^B\,,
\eea
and $\gamma_{ij}$ are given as:
\bea\label{Def:Scalar-gammas}
\gamma_{12}&=&\frac{\Delta_1+\Delta_2-h-i\nu +J}{2}\,,~ \gamma_{34}=\frac{\Delta_3+\Delta_4-h+i\nu +J}{2}\,\nn\\
\gamma_{01}&=&\frac{\Delta_{12}+h+i\nu +J}{2}\,,~ \gamma_{02}=\frac{-\Delta_{12}+h+i\nu +J}{2}\,,\nn\\
\gamma_{03}&=&\frac{\Delta_{34}+h-i\nu +J}{2}\,,~ \gamma_{04}=\frac{-\Delta_{34}+h-i\nu +J}{2}\,.
\eea
\paragraph{}
The contraction of symmetric traceless transverse tensors is implemented through the derivative $\mathcal{D}_{Z_0} \equiv \cD_{Z_0^A}$ and polarization vector $Z_0^A$, 
which is evaluated in terms of the Gegenbauer polynomial\footnote{
Here we used the following identity about the Gegenbauer polynomial $C^{(h-1)}_J (x)$:
\bea
\frac{1}{J!(h-1)_J}(X\cdot  \mathcal{D}_Z)^J (Y\cdot Z_0)^J&=& \frac{J!}{2^J (h-1)_J}(X^2 Y^2)^{\frac{J}{2}}C^{(h-1)}_J (x) \qquad \text{where}~~x=\frac{X\cdot Y}{(X^2 Y^2)^{\frac{1}{2}}}\nn\\
&=&\sum_{r=0}^{[J/2]} (-1)^r \frac{J! (J+h-1)_{-r}}{2^{2r} r! (J-2r)! }(X\cdot Y)^{J-2r}(X^2 Y^2)^{r}\,.
\eea}:
\bea
&&\frac{1}{J! (h-1)_J}(2P_{0}\cdot\cP_{12}\cdot \cD_{Z_0})^{J}(2P_0\cdot \cP_{34}\cdot Z_0)^J\nn\\
&=& \tilde{\sum_r}(-4P_0\cdot \cP_{12}\cdot \cP_{34}\cdot P_0)^{J-2r} (-4P_0\cdot \cP_{12}\cdot \cP_{12}\cdot P_0)^r(-4P_0\cdot \cP_{34}\cdot \cP_{34}\cdot P_0)^r\nn\\
&=& \tilde{\sum_r} \frac{(J-2r)!}{2^{J-2r}} P_{12}^rP_{34}^r \sum_{\sum k_{ij}=J-2r} (-1)^{k_{24}+k_{13}}
 \prod_{(ij)}\frac{P_{ij}^{k_{ij}}}{k_{ij}!} \prod_i P_{0i}^{J-r-\sum_j k_{ji}}\,,\label{Z-contraction}
\eea
where the symmetric indices $k_{ij}= k_{ji}$ comes form the expansion of the factor {$(-4P_0\cdot \cP_{12}\cdot \cP_{34}\cdot P_0)^{J-2r}$}, and $(ij)$ runs over $(13)$, $(14)$, $(23)$ and $(24)$, 
and label the all possible four-fold non-negative integer partitions of $J-2r$.
For the summation of $r$, we introduced a short-hand notation:
\bea\label{ShortSummation}
\tilde{\sum_r}=\sum^{[J/2]}_{r=0}(-1)^r\frac{J!(J+h-1)_{-r}}{2^{2r} r! (J-2r)!}\,,
\eea
{where the square parentheses $[J/2]$ is the Gauss's symbol.}
Summarizing, the boundary integration over $P_0$ now becomes:
\bea
\cI_{\nu,J}(P_i)&=&
\tilde{\sum_r} \frac{(J-2r)!}{2^{J-2r}}\frac{1}{P_{12}^{\gamma_{12}-r}P_{34}^{\gamma_{34}-r}} \\
&&\quad \times \sum_{\sum k_{ij}=J-2r} (-1)^{k_{24}+k_{13}}
 \prod_{(ij)}\frac{P_{ij}^{k_{ij}}}{k_{ij}!} 
 \int_{\partial{\rm AdS_{d+1}}} dP_0 \left(\prod_{i=1}^4 P_{0i}^{-\gamma_{0i}+J-r-\sum_j k_{ji}}\right)\,.\nn
\eea
We can now use the Symanzik star-formula\footnote{
Note here that $\sum_i\left(-\gamma_{0i}+J-r-\sum_jk_{ji}\right)=-d$. 
Therefore we can use the result for the boundary integration in Appendix \ref{App:Normal-Symanzik}. } to rewrite $\cI_{\nu, J}(P_i)$ into the desired Mellin transformation:
\bea
\cI_{\nu,J}(P_i)&=&
\tilde{\sum_r} \frac{(J-2r)!}{2^{J-2r}}\frac{1}{P_{12}^{\gamma_{12}-r}P_{34}^{\gamma_{34}-r}}  \sum_{\sum k_{ij}=J-2r} (-1)^{k_{24}+k_{13}}
 \prod_{(ij)}\frac{P_{ij}^{k_{ij}}}{k_{ij}!} \nn\\
 &\times & 
 \pi^h \left(\prod_{i=1}^4\frac{1}{\Gamma\left(\gamma_{0i}-J+r+\sum_j k_{ij}\right)} \right) \int_{-i\infty}^{i\infty}\frac{d \bar{\delta}_{12} d \bar{\delta}_{13} }{(2 \pi i)^2} \prod_{i<j}\Gamma(\bar{\delta}_{ij})P_{ij}^{-\bar{\delta}_{ij}}\,.
\eea
Here the parameters $\{\bar{\delta}_{ij}\}$ satisfy the following conditions:
\bea
\sum_{j(\neq i)} \bar{\delta}_{ij}= \gamma_{0i} -J+r +\sum_{j}k_{ij}\,,
\eea
and we can further unify the powers of $P_{ij}$ using $\delta_{ij} = \delta_{ji}$, $i \neq j$ defined as:
\be\label{delta-shift}
\delta_{12}=\bar{\delta}_{12}+\gamma_{12}-r\,,\quad 
{\delta_{34}=\bar{\delta}_{34}+\gamma_{34}-r}
\,,\quad
\delta_{(ij)}=\bar{\delta}_{(ij)}-k_{(ij)}\,,
\ee
which again satisfy the constraints {$\sum_{j (\neq i)} \delta_{ij} = \Del_i$}, $i=1,2,3,4$.
In terms of $\{\delta_{ij}\}$, $\cI_{\nu, J}(P_i)$ is expressed as:
\bea
\cI_{\nu,J}(P_i)&=& \pi^h \left(\prod_{i=1}^4\frac{1}{\Gamma(\gamma_{0i})}\right)\tilde{\sum_r} \frac{(J-2r)!}{2^{J-2r}} 
\sum_{\sum k_{ij}=J-2r} (-1)^{k_{24}+k_{13}} 
\prod _{i=1}^4\Bigl(\gamma_{0i}-J+r+\sum_jk_{ij}\Bigr)_{J-r-\sum_jk_{ij}}
\nn\\
&& \qquad \times 
\int_{-i\infty}^{i\infty}\frac{d \delta_{12} d\delta_{13} }{(2 \pi i)^2} 
\frac{\Gamma(\bar{\delta}_{12})\Gamma(\bar{\delta}_{34})}{\Gamma(\delta_{12})\Gamma(\delta_{34})}  
 \prod_{(ij)}\frac{(\delta_{ij})_{k_{ij}}}{k_{ij}!}
 \prod_{i<j} \Gamma(\delta_{ij}) P_{ij}^{-\delta_{ij}}\,,
\eea
where the additional Pochhammer symbols arise from the shifts in \eqref{delta-shift}.
If we now identify the Mellin momenta $s$ and $t$ as:
\bea
s=\Delta_{34}-2 \delta_{13}\,, \quad t = \Delta_1 +\Delta_2 -2 \delta_{12}\,,
\eea
then the relation between $\delta_{ij}$ and $(s,t)$ are given in \eqref{Mellin-deltaij}, and we obtain the following form:
\bea
\cI_{\nu,J}(P_i)&=&\pi^h \left(\prod_{i=1}^4\frac{1}{\Gamma(\gamma_{0i})}\right) \int_{-i\infty}^{i\infty}\frac{d sdt }{(4 \pi i)^2}
\frac{\Gamma\left(\frac{h\pm i\nu -J -t}{2}\right)}{\Gamma\left(\frac{\Delta_1+\Delta_2 -t}{2}\right)\Gamma\left(\frac{\Delta_3+\Delta_4 -t}{2}\right)}
 \tilde{P}_{\nu,J}(s,t) \prod_{i<j} \Gamma(\delta_{ij}) P_{ij}^{-\delta_{ij}}\,,\nn\\
 \label{Scalar-Integral1}
\eea
where $\tilde{P}_{\nu,J}(s,t)$ is the following polynomial:
\bea\label{Def:tildeP}
\tilde{P}_{\nu,J}(s,t)&=&\sum^{[J/2]}_{r=0}(-1)^r\frac{J!(J+h-1)_{-r}}{2^{J-2r} r! } \left(\frac{h\pm i\nu-J-t}{2}\right)_r\nn\\
&\times&
 \sum_{\sum k_{ij}=J-2r} (-1)^{k_{24}+k_{13}}  \prod_{(ij)}\frac{(\delta_{ij})_{k_{ij}}}{k_{ij}!}
 \prod _{i=1}^4\Bigl(\gamma_{0i}-J+r+\sum_jk_{ij}\Bigr)_{J-r-\sum_jk_{ij}}\,.
\eea
Notice that we have absorbed the shift by $r$ in $\bar{\delta}_{12}$ and $\bar{\delta}_{34}$ in \eqref{delta-shift} by introducing additional $r$-dependent Pochhammer symbols in \eqref{Def:tildeP}.
{This polynomial}
differs from the Mack polynomial $P_{\nu, J}(s,t)$ given in \cite{ConformalRT} by an overall factor\footnote{Notice also that up to $(-1)^J$ factor,  $\tilde{P}_{\nu, J}(s, t)$ here is identical as the Mack polynomial used in \cite{Gopakumar2016-1}, \cite{Gopakumar2016-2}. In making this statement, we have also used a useful identity \eqref{Id-Mack} proven in the appendix \ref{App:Id-Mack}.}:
\be\label{Conversion}
P_{\nu,J}(s,t)= \frac{2^{2J}}{(h\pm i\nu-1)_J}\tilde{P}_{\nu,J}(s,t)\,.
\ee
This demonstrates that up to the overall factors above which gives the spurious poles, Mack polynomial appearing in the scalar Mellin amplitude \eqref{MellinPartial}
purely comes from the kinematic integral \eqref{Def:ScalarKI}, which in turns can be fixed by conformal symmetries alone. 
The remaining dynamical coefficients in \eqref{4pt-Witten2} are evaluated as:
\bea
&&\frac{2\nu^2}{\nu^2 +(\Delta-h)^2} b(\Del_1, \Del_2,h+i\nu , J) b(\Del_3, \Del_4, h-i\nu, J)\nn\\
&&= \frac{2^{2J-3}}{\nu^2 +(\Delta-h)^2}\left(\prod_{i=1}^4\frac{\cC_{\Delta_i,0}}{\Gamma(\Delta_i)}\Gamma(\gamma_{0i})\right)\frac{\Gamma\left(\frac{\Delta_1+\Delta_2+J-h\pm i\nu}{2}\right)\Gamma\left(\frac{\Delta_3+\Delta_4+J-h\pm i\nu}{2}\right)}{\Gamma(\pm i\nu)(h\pm i\nu-1)_J},\label{RemainingFactors}
\eea
which is to be combined with the pre-factor multiplying $\tilde{P}_{\nu, J} (s, t)$. 
Finally we can repackage \eqref{4pt-Witten2} as 
\bea
W^{\text{4pt}}_{J} (P_i) &=&
\frac{\pi^h g_{\Phi_1\Phi_2\Xi}g_{\Phi_3\Phi_4\Xi}  }{2} 
\left(\prod_{i=1}^4\frac{{\cC_{\Delta_i,0}}}{\Gamma(\Delta_i)}\right) \int^{+\infty}_{-\infty} d\nu
\frac{1}{\nu^2+(\Del-h)^2} \\
&&\qquad\qquad\qquad\qquad\qquad \times \int_{-i\infty}^{i\infty}\frac{d sdt }{(4 \pi i)^2} \omega_{\nu,J}(t) P_{\nu,J}(s,t) \prod_{i<j} \Gamma(\delta_{ij})P_{ij}^{-\delta_{ij}}\nn\\
&=&
 \int_{-i\infty}^{i\infty}\frac{d sdt }{(4 \pi i)^2} \int^{+\infty}_{-\infty} d\nu\, b^{(\text{scalar})}_J(\nu)  \cM_{\nu,J}(s,t) \prod_{i<j} \Gamma(\delta_{ij})P_{ij}^{-\delta_{ij}}\,,
 \label{4pt-Witten3}
\eea
where $\omega_{\nu,J}(t)$ is given in \eqref{Omega-t} and if we identify $W^{\text{4pt}}_J(P_i)$ with the four point scalar correlation function \eqref{MellinTrans-4ptScalar},
this requires us to set the spectral function as:
\be \label{Def:func-b}
b^{(\text{scalar})}_J(\nu)= \frac{\pi^h g_{\Phi_1\Phi_2\Xi}g_{\Phi_3\Phi_4\Xi}  }{2} 
\left(\prod_{i=1}^4\frac{{\cC_{\Delta_i,0}}}{\Gamma(\Delta_i)}\right) \frac{1}{\nu^2+(\Del-h)^2}.
\ee
Before moving on, it is worthwhile to pause here to comment on the general structure of scalar four point Witten diagram/correlation function leading to the Mellin amplitude \eqref{MellinAmp}, as it will be useful for us to consider the spinning generalization next.
The singularities at $\pm i\nu=t+J-h + 2m$ are purely kinematical and arise from {the $\Gamma$-functions in fusing two copies of the box basis in \eqref{Scalar-Integral1}}. As these singularities collide with the poles from $\pm i\nu = \Del-h$ in \eqref{RemainingFactors} arising from the cutting representation of the bulk-bulk propagator \eqref{Cutting Identity}, we can obtain the expansion \eqref{MellinExpansion} around $t= \Del-J+2m$. We will see this mechanism of colliding the singularities of the cutting representation of bulk-bulk propagator and kinematical integral for generating the desired t-channel singularities also applies to the spinning Mellin amplitudes.
\paragraph{}
Moreover as can be seen from \eqref{RemainingFactors}, the contributions from the double trace operators to the full correlation function are contained in the dynamical pre-factors $b(\Del_1, \Del_2, h+i\nu, J)$ and $b(\Del_3, \Del_4, h-i\nu, J)$ multiplying the box tensor basis to yield the three point Witten diagrams.  
Naively, we can use the same pole colliding mechanism to generate the $t$-channel singularities corresponding double trace operators in the scalar Mellin amplitude, 
however as we parameterize it, these are precisely canceled by the zeroes coming from the $\Gamma(\frac{\Del_1+\Del_2-t}{2}) \Gamma(\frac{\Del_3+\Del_4-t}{2})$ in the denominator.
In the spinning generalization, the relevant three point Witten diagrams are expressed in terms of linear combination of box tensor basis, different choices of three point interacting vertices generate different dynamical pre-factors, and these only affect the spectrum of the double trace operators. 
We will see that using appropriate identification of Mellin momenta and parameterization, we can ensure the resultant spinning Mellin amplitude is free of double trace operator singularities.
\paragraph{}
Another side comment is that, the coefficients $\hb(\Del_1, \Del_2, h+i\nu, J)$ and $\hb(\Del_3, \Del_4, h-i\nu, J)$ 
coming form geodesic diagrams do not contain any poles associated with the double trace operators as in \eqref{Def:hbJnu}.
This means the fact that if we consider a four point geodesic diagram which is constructed by gluing two three point geodesic diagrams,
the Mellin amplitude has only zeros for the double trace operators, and the four point geodesic diagram has only the single trace contribution.
More concretely, from the geodesic diagram, the following functions are obtained instead of $b_J(\nu)$ and $\omega_{\nu,J}(t)$ in \eqref{Def:func-b} and \eqref{Omega-t}:  
\bea
b_J^{\text{GWD}}(\nu)&=&\frac{g_{\Phi_1\Phi_2\Xi}g_{\Phi_3\Phi_4\Xi}  }{2\pi^h } 
\left(\prod_{i=1}^4\cC_{\Delta_i,0}\right) (\Delta_2)_J(\Delta_4)_J \frac{1}{\nu^2+(\Del-h)^2}\,,\nn\\
\omega_{\nu,J}^{\text{GWD}}(t)&=&\frac{\Gamma\left(\frac{h\pm i\nu -J-t}{2}\right)}{8\pi \, \Gamma(\pm i\nu) 
\Gamma\left(\frac{\Delta_1+\Delta_2-t}{2}\right)
\Gamma\left(\frac{\Delta_3+\Delta_4-t}{2}\right)}\,.
\eea
In this sense, the geodesic diagrams are more fundamental elements, and actually, 
expanding the $\Gamma$-functions including the double trace poles,
we can decompose normal four point exchange Witten diagrams into a summation of geodesic diagrams with the single and double trace operators exchanges \cite{SpinningGWD}.

\section{Spinning Mellin Amplitudes from Witten Diagrams}\label{Sec:Spinning Mellin}
\subsection{Spinning Three Point Functions in CFT and AdS}\label{Sec:SpinningThreePoint}
\paragraph{}
Let us now consider the Mellin representation of four point correlation function involving four external primary operators with integer spins $\cO_{\Del_i, l_i}(P_i, Z_i)$, $i=1,2,3,4$. 
Generally, in addition to the symmetric traceless primary tensor operators as in the scalar case, the OPE between a pair of spinning primary operators $\{\cO_{\Del_i, l_i}(P_i, Z_i)\}$ can also contain mixed tensor primary operators, these have been considered in several recent works, e.g. \cite{SeedBlock1, MixedTensor1, GeneralBlock1, GeneralBlock2, GeneralBlock3}.  
If we however restrict to only the symmetric traceless exchange primary operators, this situation is somewhat simplified and was considered several years ago in \cite{SpinningBlock0}, \cite{SpinningBlock1}, 
we briefly review the essential details here before considering their corresponding Mellin representation.
\paragraph{}
We start again with the general three point functions for primary operators with integer spins which can be expressed in terms of linear combinations of the following box tensor basis:
\bea\label{Spinning3ptFunction}
&&\langle \cO_{\Del_1, l_1}(P_1, Z_1)\cO_{\Del_2, l_2}(P_2, Z_2)\cO_{\Del_0, l_0}(P_0, Z_0)\rangle_{\rm CFT} \nn\\
&&\qquad\qquad\qquad\qquad= \sum_{n_{12}, n_{10}, n_{20} \ge 0} \lam_{n_{12}, n_{10}, n_{20}} 
\begin{bmatrix}
				\Delta _1 &\Delta _2&  \Del_0    \\[0.05em]
				l_1 & l_2           & l_0 \\[0.05em]
				n_{20}         &     n_{10}     & n_{12}  
\end{bmatrix}
+\cO(Z_i^2, Z_i\cdot P_i)\nn
\eea
\be\label{CFT-BoxBasis}
\begin{bmatrix}
				\Delta _1 &\Delta _2&  \Del_0    \\[0.05em]
				l_1 & l_2           & l_0 \\[0.05em]
				n_{20}         &     n_{10}     & n_{12}  
\end{bmatrix}
= \frac{V_{1,20}^{m_1} V_{2, 01}^{m_2} V_{0, 12}^{m_0}H_{10}^{n_{10}}H_{20}^{n_{20}}H_{12}^{n_{12}}}{(P_{10})^{\frac{1}{2}(\tau_1+\tau_0-\tau_2)} (P_{20})^{\frac{1}{2}(\tau_2+\tau_0-\tau_1)} (P_{12})^{\frac{1}{2}(\tau_1+\tau_2-\tau_0)} }	
\ee
where $\tau_i = \Del_i+l_i$, and $\{\lambda_{n_{10}, n_{20}, n_{12}}\}$ are the expansion coefficients associated with the independent tensor structures labeled by\footnote{In \eqref{Spinning3ptFunction} $\cO(Z_i^2, Z_i\cdot P_i)$ denote the terms which are not linearly independent, and corresponding to the same three point function.
{For symmetric traceless transverse tensor fields, the polarization vectors $Z_i$ should satisfy $Z_i^2=Z_i\cdot P_i=0$,
therefore we can ignore these terms.}}:
\be\label{TensorStructures}
{V_{i, jk} = \frac{-2 P_j \cdot {\rm C}_i \cdot P_k}{P_{jk}} = \frac{{\rm Tr}(\rC_i {\mathcal{P}}_{jk})}{P_{jk}}, \quad H_{ij} = {\rm Tr}(\rC_i \rC_j).}
\ee
Here we have introduced the following rank-two anti-symmetric tensors 
\be\label{AntisymTensor}
\rC_{i}^{AB} = Z_i^A P_i^B - P_i^A Z_i^B, \quad  \cP_{jk}^{AB} = P_j^A P_k^B-P_j^B P_k^A, \quad  i, j, k=0,1,2 
\ee
which ensure the tensor structures \eqref{TensorStructures} are invariant under the shift {$Z_i^A \to Z_i^A + P_i^A$}.
Moreover, the triplet of non-negative integers $(n_{10}, n_{20}, n_{12})$ need to satisfy the following partition constraints:
\be\label{Constraints}
m_1 = l_1 - n_{10} - n_{12} \ge 0, \quad m_2 = l_2 - n_{20}-n_{12} \ge 0, \quad m_0 = l_0 - n_{10} - n_{20} \ge 0\,.
\ee
For given $l_1\leq l_2\leq l_0$ with $p \equiv {\rm max}(0, l_1+l_2=l_0)$, we have
\be\label{Number-box}
N(l_1, l_2, l_0) = \frac{(l_1+1)(l_1+2)(3l_2-l_1+3)}{6} - \frac{p(p+2)(2p+5)}{24}-\frac{1-(-1)^p}{16}
\ee
independent box {tensor structures} listed in \eqref{CFT-BoxBasis}.
\paragraph{}
Holographically, we can again compute the spinning three point functions by applying Witten diagram technique to three AdS tensor fields with integer spins.
One important difference from the scalar-scalar-tensor computation reviewed earlier however is that, instead of a unique interaction vertex \eqref{SST-3ptVertex},
there can in general be multiple number of independent interaction vertices contributing, each contains different number of spacetime derivatives and index contractions. 
This is in accordance with the existence of multiple independent tensor box basis in \eqref{Spinning3ptFunction} for expressing the CFT three point functions.
\paragraph{}
Recently there have been remarkable progress in parameterizing all possible three point interaction vertices for AdS symmetric {traceless} tensor fields with integer spins, 
and the computation of their corresponding contact Witten diagrams \cite{3pt-Coupling}, \cite{SpinningWitten}.
\footnote{
The Mellin representations for $n$-point contact Witten diagrams with spins can be also obtained by using the generalized Symanzik formula basically, however, the results usually would be complicated.
}
In particular, the number of independent interaction vertices in the AdS space precisely equals to the number of independent box tensor basis given in \eqref{Number-box}, 
therefore expressing the resultant three point Witten diagram for a given interaction vertex to the box tensor basis essentially becomes a diagonalization problem.
In its state of art form \cite{SpinningWitten}, the resultant expression  can be summarized into the following general structure:
\be\langle \cO_{\Del_1, l_1}(P_1, Z_1)\cO_{\Del_2, l_2}(P_2, Z_2)\cO_{\Del_0, l_0}(P_0, Z_0)\rangle_{\rm WD} =\sum_{\{{\bk}\}} \lam^{{\bk}}_{\bl} \cA^{{\bk}}_{{\btau}, {\bl}}(P_i, Z_i), \quad i = 0,1,2
\label{3pt-spinning-generic}
\ee
where $\cA^{{\bk}}_{{\btau}, {\bl}}(P_i, Z_i)$ denotes the contribution from a specific interaction vertex $\cV^{\bk}_{\bl}$ labeled by triplet of non-negative integers $\bk=(k_1, k_2, k_0)$, their scaling dimensions
$\btau =(\tau_1, \tau_2, \tau_0)$ and spins $\bl =(l_1, l_2, l_0)$.
The corresponding contribution $\cA^{{\bk}}_{\btau, \bl}$, after the integration over entire AdS space, can be expressed in turns as the linear combination of the box tensor basis 
\eqref{CFT-BoxBasis}:
\bea\label{A-contribution}
 \cA^{{\bk}}_{{\btau}, {\bl}}(P_i, Z_i) &=& \bB(\bl, \bk;\btau) \sum_{\{\bn\}} f_{\bn, \bl}^{\bk} \left(\delta_{(i-1)(i+1)}\right) \begin{bmatrix}
~\Delta_1~ &~ \Delta_2 ~&~ \Delta_0~\\
l_1 & l_2 & l_0\\
n_{20}& n_{10} & n_{12}
\end{bmatrix}\,,\\
\delta_{(i-1)(i+1)} &=& \frac{\tau_{i+1}+\tau_{i-1}-\tau_i}{2},\qquad i\cong i+3.\nn
\eea
Here we have used triplet of non-negative integers $\bn=(n_{10}, n_{20}, n_{12})$ to denote all possible partitions which depend on spins $\bl$, 
and they satisfy the partition constraints \eqref{Constraints}.
The expansion coefficients $f_{\bn, \bl}^{\bk} \left(\delta_{(i-1)(i+1)}\right) $ only depend 
on the scaling dimensions through the combinations $\delta_{(i-1)(i+1)}$, where one of the $\tau_i, \tau_{i\pm 1}$ will be promoted to spectral parameter later.
In essence, the structure of the spinning three point Witten diagram summarized in \eqref{3pt-spinning-generic} and \eqref{A-contribution} 
are generalization of the scalar-scalar-tensor three point Witten diagram \eqref{Def:SST-3ptGWD}, where only unique box tensor {structure} is needed.
\paragraph{}
When we use the cutting identity \eqref{Cutting Identity} to fuse two
spinning three point Witten diagrams and proceed to obtain the spinning Mellin amplitude,
the combined dynamical pre-factors $\bB(\bl, \bk; \btau)f_{\bn, \bl}^{\bk} \left(\delta_{(i-1)(i+1)}\right) $ 
multiplying the individual box tensor basis play the same role as the factor 
\eqref{Def:bJnu}.
in the derivation of scalar Mellin amplitude, 
whose singularities determine the spectrum of double trace operators as we integrate over the spectral parameter $\nu$. 
In deducing the Mellin representation of the spinning correlation function from spinning Witten diagrams, 
we should again have clear separation between the contributions from single trace and double trace operators.
We will show that these singularities corresponding to the double trace operators can be separated from the spinning Mellin amplitude as in the scalar case.
{Another basis for the three point interaction vertices is also possible \cite{3pt-Coupling}, and they are related through an appropriate mixing matrix.}
\paragraph{}
Putting various pieces together, now for a given pair of three point interaction vertices and their corresponding $\cA_{\btau_L, \bl_L}^{\bk_L}$ and $\cA_{\btau_R, \bl_R}^{\bk_R}$,
where the parameters are:
\bea
&&\btau_L = (\tau_1, \tau_2, h+i\nu+J), \quad \bk_L = (k_1, k_2, k_0), \quad \bl_L =(l_1, l_2, J),\nn\\
&&\btau_R = (\tau_3, \tau_4, h-i\nu+J), \quad \bk_R = (k_3, k_4, \tilde{k}_0), \quad \bl_R =(l_3, l_4, J).
\eea
We can now parameterize their combined contributions to the four point spinning Witten diagram schematically as:
\bea\label{four point Witten}
W^{\text{4pt}}_{\Delta,J}({\bf k}_L,{\bf k}_R) &=&g_{\Xi_1\Xi_2\Xi_0}g_{\Xi_3\Xi_4\tilde{\Xi}_0}
\int^{+\infty}_{-\infty}\frac{d \nu}{2 \pi} \frac{2\nu^2}{\nu^2+(\Delta-h)^2} \\
&&\qquad \qquad \qquad \times \sum_{\{{\bf n}_L,{\bf n}_R\}}
 \mathbf{b}({\bf k}_L,{\bf n}_L) \mathbf{b}({\bf k}_R,{\bf n}_R) ~\cI^{({\bf n}_L,{\bf n}_R)}_{\nu, J}(P_i, Z_i).\nn
\eea
Here {$\mathbf{b}(\bk_{L, R},{\bf n}_{L, R})$} are the the combined dynamical pre-factors, i.e. $\bB(\bl, \bk;\btau)$ and $f_{\bn, \bl}^{\bk} \left(\delta_{(i-1)(i+1)}\right)$, 
multiplying the individual box tensor basis involved in \eqref{A-contribution}, 
they encode the specific choice of basis for the three point interaction vertices. 
For each pair of triplets of integers $(\bn_L, \bn_R)$, their corresponding box tensor basis fused into the kinematic integral $\cI^{(\bn_L, \bn_R)}_{\nu, J}(P_i, Z_i)$.
As we reviewed in the previous section, evaluating the kinematic integral $\cI_{\nu, J}(P_i)$ using Symanzik star-formula is a crucial step for obtaining the scalar Mellin partial amplitude.
Here we would like to generalize the Symanzik star-formula to the kinematic integral $\cI^{(\bn_L, \bn_R)}_{\nu, J}(P_i, Z_i)$ for spinning four point correlation functions, 
and we will argue that this should be regarded as the natural kinematic basis for constructing the full spinning Mellin amplitude. 


\subsection{Evaluation of the Spinning Kinematical Integral}\label{Sec:Spinning Integral}
\paragraph{}
In this section, for a given pair of interaction vertices in \eqref{four point Witten} labeled by $(\bk_L,\bk_R)$,
we proceed to evaluate  the spinning kinematical integral $\cI^{(\bn_L, \bn_R)}_{\nu, J} (P_i, Z_i)$ which includes external polarization vectors $\{Z_i\}$:
\bea \label{general integral 0}
\cI^{({\bf n}_L,{\bf n}_R)}_{\nu, J}(P_i, Z_i)
\equiv\frac{1}{J!\, (h-1)_J}\int_{\partial}dP_0 
\begin{bmatrix}
~\Delta_1~ &~ \Delta_2 ~&~ h+i\nu~\\
l_1 & l_2 & J\\
n_{20}& n_{01} & n_{12}
\end{bmatrix}
\cdot
\begin{bmatrix}
~\Delta_3~ &~ \Delta_4 ~&~ h-i\nu~\\
l_3 & l_4 & J\\
n_{40}& n_{03} & n_{34}
\end{bmatrix}\,.
\eea
This integration can be evaluated and expressed into Mellin representation by using the generalization of 
Symanzik star-formula given in Appendix \ref{App:Generalized Symanzik}, and 
we relegate most of the calculation details and some definitions of notations to Appendix \ref{App:CI Spinning}. 
Let us first state the final result:
\bea\label{general integral2}
\cI^{({\bf n}_L,{\bf n}_R)}_{\nu, J}(P_i, Z_i)
&=&
\pi^h \left(\prod_{i=1}^4\frac{m_i!}{\Gamma (\tilde{\gamma}_{0i})}\right)
\sum_{\{a_{ij},b_{ij}\}} \int_{-i\infty}^{i\infty} \frac{d\delta_{12}d\delta_{13}}{(2 \pi i)^2}\,\\
&&\qquad \times 
 \tilde{P}_{\nu,J}^{({\bf n}_L,{\bf n}_R)}(\delta_{ij},a_{ij},b_{ij})
\frac{\Gamma(\hat{\delta}_{12})\Gamma(\hat{\delta}_{34})}{\Gamma(\delta_{12})\Gamma(\delta_{34})}
\prod_{i\neq j} \frac{\tilde{V}_{ij}^{a_{ij}}}{a_{ij}!}
\prod_{i<j}\frac{\tilde{H}_{ij}^{b_{ij}}}{b_{ij}!} \Gamma(\delta_{ij})P_{ij}^{-\delta_{ij}}\,,\nn
\eea
and explain various quantities and definitions in turns.
Here the Mellin variables $\delta_{ij}$ satisfy the modified constraints:
\be\label{constraint2}
\sum_{j(\neq i)}\delta_{ij}=\Delta_i - l_i\equiv \tilde{\tau}_i\,,
\ee
where {we have introduced the twist parameters $\tilde{\tau}_i$} and we can solve the constraint \eqref{constraint2} as in \eqref{Mellin-deltaij} with $\Del_i$ replaced by $\tilde{\tau}_i$.
This also leads us to the natural identifications of Mellin momenta for the spinning Mellin amplitude as:
\bea\label{Spinningst}
s=\tilde{\tau}_3-\tilde{\tau}_4-2 \delta_{13}\,, \quad t = \tilde{\tau}_1 +\tilde{\tau}_2 -2 \delta_{12}\,.
\eea
The arguments $\hat{\delta}_{12}$ and $\hat{\delta}_{34}$ in the explicit $\Gamma$-functions in \eqref{general integral2} are:
\bea\label{hat delta}
\hat{\delta}_{12} &=& \delta_{12}-\frac{\tilde{\tau}_1+\tilde{\tau}_2-\tilde{\tau}_0^+}{2}\,,\qquad 
\hat{\delta}_{34} = \delta_{34}-\frac{\tilde{\tau}_3+\tilde{\tau}_4-\tilde{\tau}_0^-}{2}\,,
\eea
where 
$\tilde{\tau}^\pm_0\equiv h \pm i\nu -J$\,.
In terms of the Mellin momenta $t$, $\hat{\delta}_{12}$ and $\hat{\delta}_{34}$ can be expressed as:
\bea\label{hat delta2}
\hat{\delta}_{12} &=&\frac{h+i\nu-J-t}{2}\,,\qquad 
\hat{\delta}_{34} = \frac{h-i\nu-J-t}{2}\,.
\eea
In \eqref{general integral2}, we have introduced the alternative basis for independent tensor structures $\{\tilde{V}_{ij}\}$ and $\{\tilde{H}_{ij}\}$:
\bea \label{VH-tilde basis}
\tilde{V}_{ij}\equiv \frac{-2 \bar{v}_i \cdot P_j}{P_{ij}}\,, \quad
\tilde{H}_{ij} \equiv \frac{-2 \bar{v}_i \cdot \bar{v}_j}{P_{ij}}, \quad i, j =1,2,3,4.
\eea
Here $\{\bar{v}_i^A$\} are composite vectors which are constructed from antisymmetric tensor $\rC_i^{AB}$, and they are defined as:
\bea\label{def:bar v}
\bar{v}_1^A\equiv \frac{(-2P_2\cdot \rC_1)^A}{P_{12}}\,,~ \bar{v}_2^A\equiv \frac{(2P_1\cdot \rC_2)^A}{P_{12}}\,,~
 \bar{v}_3^A\equiv \frac{(-2P_4\cdot \rC_3)^A}{P_{34}}\,,~ \bar{v}_4^A\equiv \frac{(2P_3\cdot \rC_4)^A}{P_{34}}\,.
\eea
The list of the non-vanish products among $P_i$ and $\bar{v}_i$ is given in \eqref{list P dot v} and \eqref{list v dot v}, 
which consist of combinations of $\{V_{i, jk}\}$ and $\{H_{ij}\}$ given in \eqref{TensorStructures}.
In particular $\tV_{ij}$ is linear in $Z_i$ and $\tH_{ij}$ is quadratic in $Z_i$ and $Z_j$, and we still preserve the transverse condition.
The two sets of non-negative integers $\{a_{ij}\}$ and $\{b_{ij}\}$ labeling the powers of $\tV_{ij}$ and $\tH_{ij}$ need to satisfy the constraints:
\be\label{constraint-ab}
\sum_{j(\neq i)}(b_{ij}+a_{ij})=l_i\,,
\ee
this can be understood that we can only have total $l_i$ of $Z_i$ polarization vectors in the final expression.
The $\{a_{ij}\}$ and $\{b_{ij}\}$ in the summation of \eqref{general integral2} take values when they satisfy the constraint \eqref{constraint-ab}.
Notice that $b_{ij}$ are symmetric in the two indices, but $a_{ij}$ are not, and they can be regarded as the discrete Mellin variables 
which incorporate the discrete spin degrees of freedom encoded in the invariant tensor structures $\{\tilde{H}_{ij}, \tilde{V}_{ij}\}$.
\paragraph{}
Now to count the number of independent parameters, both continuous and discrete, there are two types of constraints: \eqref{constraint2} for $\delta_{ij}$ and \eqref{constraint-ab} for $a_{ij}$ and $b_{ij}$\,.
There are six continuous variables $\delta_{ij}$ initially,  the constraint \eqref{constraint2} eliminate them to only two independent ones, same number as the independent cross ratios.
Similarly, due to the four constraints \eqref{constraint-ab}, starting with twelve $a_{ij}$ and six $b_{ij}$, we are left with fourteen independent discrete variables.
This number precisely corresponds to the number of independent elements of tensor structures, explicitly we have eight $\tilde{V}_{ij}$'s and six $\tilde{H}_{ij}$'s as in 
from \eqref{list P dot v} to \eqref{Def:v and h}. 
\paragraph{}
It is also interesting to pause here and consider the the possible interpretations of these remaining variables in the flat space limit.
For the continuous Mellin variable $\delta_{ij}$, we can again regard them as bilinear of scattering momenta $\{p_i\}$ in the flat space through the identification: $\delta_{ij}=p_i\cdot p_j$ and $p_i^2=-\tilde{\tau}_i$.
For the flat space limit of independent discrete $\{a_{ij}, b_{ij}\}$, we can count the number of independent possible tensor structures arising from the four point spinning scattering amplitudes.
Consider the fields with spins in the flat space following \cite{FactMellin}, 
we can also construct the corresponding elements of tensor structures as products of the scattering momenta $\{p_i\}$ and polarization vectors $\{\epsilon_i\}$, $i=1,2,3,4$ for each spinning field.
There are six $\epsilon_i\cdot \epsilon_j$ and eight $\epsilon_i\cdot p_j$ because inner products like $\epsilon_i\cdot \epsilon_i = 0$ or $p_i\cdot \epsilon_i = 0$.
The number of the independent products are the same as the discrete Mellin variables $\{a_{ij}\}$ and $\{b_{ij}\}$, which indicate
they may play the similar role of enumerating the independent tensor structures even in the flat space.
\paragraph{}
Finally in \eqref{general integral2}, we have introduced the following polynomial:
\bea
&& \tilde{P}_{\nu,J}^{({\bf n}_L,{\bf n}_R)}(\delta_{ij},a_{ij},b_{ij})\equiv \tilde{\sum_{r,\beta, k}}\frac{1}{(-2)^{\sum_i m_i +n_{12}+n_{34}}}  \prod_i (\tilde{\gamma}_{0i}-\kappa_i)_{\kappa_i}
(m_i+1)_{\bar{\kappa}_{\bar{i}}}
\nn\\
&&\quad \times
{
(\hat{\delta}_{12})_{d_{12}} 
(\hat{\delta}_{34})_{d_{34}}
}
\frac{b_{12}!}{(b_{12}-n_{12}-\bar{k}^{\{1\bar{1}\bar{2}2\}})!}
\frac{b_{34}!}{(b_{34}-n_{34}-\bar{k}^{\{3\bar{3}\bar{4}4\}})!}
 \nn\\
&&\quad \times 
\prod_{(ij)} \frac{(\delta_{ij})_{d_{ij}+\sum_{\{A\}} k_{ij}^{\{A\}}} }{\prod_{\{A\}} k_{ij}^{\{A\}}!} 
\prod_{(\bar{i}j)} \frac{(a_{ij}-\sum_{\{A\}}k_{\bar{i}j}^{\{A\}} +1)_{\sum_{\{A\}} k_{\bar{i}j}^{\{A\}}} }{\prod_{\{A\}} k_{\bar{i}j}^{\{A\}}!} 
\prod_{(\bar{i}\bar{j})}\frac{(b_{ij}-\sum_{\{A\}}k_{\bar{i}\bar{j}}^{\{A\}} +1)_{\sum_{\{A\}} k_{\bar{i}\bar{j}}^{\{A\}}} }{\prod_{\{A\}} k_{\bar{i}\bar{j}}^{\{A\}}!} \,,
\nn\\
\eea
where the definition of the summation $\tilde{\sum}_{r,\beta,k}$ is given in \eqref{r beta k sum}, 
$\{\tilde{\gamma}_{0i}\}$ are given in \eqref{def:tilde gamma}, and $\kappa_i$ and $\bar{\kappa}_{\bar{i}}$ are defined in \eqref{def:kappa i} as summations of integers $\beta^{\{...\}}$ and $k_{\alpha\beta}^{\{...\}}$ which come from the polynomial expansions.
{$d_{ij}$ are non-negative integers defined in \eqref{def:d} and \eqref{shift:tilde delta}.}
Note that when all external spins $l_i=0$, this polynomial becomes $\tilde{P}_{\nu,J}$ in \eqref{Def:tildeP}. 
In this sense, we shall call $ \tilde{P}_{\nu,J}^{({\bf n}_L,{\bf n}_R)}(\delta_{ij},a_{ij},b_{ij})$ ``{\it generalized Mack polynomial}'', 
{which is purely kinematical and can be regarded as the natural polynomial basis for expressing spinning conformal partial waves in Mellin space.}
\paragraph{}
Combining various pieces, now we can define the Mellin representation for the contribution of interaction vertices labeled by $(\bk_L, \bk_R)$ to the four point spinning Witten diagram.
Substituting \eqref{general integral2} into \eqref{four point Witten}, we obtain the following:
\bea\label{Spinning4pt-Mellin}
&&\!\!\!\!W^{\text{4pt}}_{\Delta,J}({\bf k}_L,{\bf k}_R)=\\
&&\!\!\!\!\sum_{\{{\bf n}_L,{\bf n}_R\}}\sum_{\{a_{ij},b_{ij}\}}\, \int^{i\infty}_{-i\infty} \frac{ds\, dt}{(4\pi i)^2}
\int^\infty_{-\infty}\! d\nu ~ b^{({\bf n}_L,{\bf n}_R)}_J(\nu)
\cM_{\nu,J}^{({\bf n}_L,{\bf n}_R)}(s,t;a_{ij},b_{ij}) \prod_{i\neq j} \frac{\tilde{V}_{ij}^{a_{ij}}}{a_{ij}!}
\prod_{i<j}\frac{\tilde{H}_{ij}^{b_{ij}}}{b_{ij}!}\prod_{i< j} \frac{\Gamma(\delta_{ij})}{P_{ij}^{\delta_{ij}}}\,,\nn
\eea
where we define the spectrum function $b^{({\bf n}_L,{\bf n}_R)}_J(\nu)$ and the spinning Mellin amplitude
$\cM_{\nu,J}^{({\bf n}_L,{\bf n}_R)}(s,t;a_{ij},b_{ij})$ as:
\bea\label{SpinningSpecFunc}
b^{({\bf n}_L,{\bf n}_R)}_J(\nu)
&\equiv&\frac{\pi^h g_{\Xi_1\Xi_2\Xi_0}g_{\Xi_3\Xi_4\tilde{\Xi}_0}}{2}
\left(\prod_{i=1}^4\frac{m_i!}{\Gamma(\tau_i)}\cC_{\Delta_i,l_i}\right) \frac{1}{(h-\Delta)^2+\nu^2}\,,
\\
\label{SpinningMellinAmp}
\cM_{\nu,J}^{({\bf n}_L,{\bf n}_R)}(s,t;a_{ij},b_{ij})
&\equiv&
\frac{2 \nu^2}{\pi}\left(\prod_{i=1}^4\frac{\Gamma(\tau_i)}{\Gamma(\tilde{\gamma_{0i}})\cC_{\Delta_i,l_i}}\right)\\
&&\qquad \times \mathbf{b}({\bf k}_L,{\bf n}_L)  
\mathbf{b}({\bf k}_R,{\bf n}_R)  
\frac{\Gamma(\hat{\delta}_{12})\Gamma(\hat{\delta}_{34})}{\Gamma(\delta_{12})\Gamma(\delta_{34})}
\tilde{P}_{\nu,J}^{({\bf n}_L,{\bf n}_R)}(\delta_{ij},a_{ij},b_{ij})\,.\nn
\eea
We can view the Mellin representation of the spinning four point function \eqref{Spinning4pt-Mellin} 
as combining the continuous integral and discrete transformations, where we have rewritten the factorials 
$a_{ij}!$ and $b_{ij}!$ into $\Gamma$-functions to make them on parallel footing with the continuous Mellin variables $\del_{ij}$.
These variables need to satisfy the constraints \eqref{constraint2} and \eqref{constraint-ab},
the last three products of Gamma functions in \eqref{Spinning4pt-Mellin} form universal transformation kernel for given external twists $\{\ttau_i\}$ and spins $\{l_i\}$.
All the information about specific choice of interaction vertices encoded in the dynamical factors $\mathbf{b}({\bf k}_L,{\bf n}_L)  
\mathbf{b}({\bf k}_R,{\bf n}_R)$, we can for example use the explicit expressions for these factors given in \cite{3pt-Coupling} and \cite{SpinningWitten} to compute the complete spinning Mellin amplitude \eqref{SpinningMellinAmp}. 
\paragraph{}
Here we would like to argue that using the identification of Mellin momenta \eqref{hat delta} and \eqref{hat delta2},
the spinning Mellin amplitude \eqref{SpinningMellinAmp} itself again does not contain any singularities corresponding to the double trace operators as in the scalar case. 
As discussed earlier, the dynamical pre-factors of the three point spinning Witten diagram $\mathbf{b}({\bf k}_L,{\bf n}_L)$ and $\mathbf{b}({\bf k}_R,{\bf n}_R)$
include poles in the $\nu$-plane corresponding to the double trace operators. Upon collision with the poles in $\Gamma(\hat{\delta}_{12})\Gamma(\hat{\delta}_{34}) = \Gamma\left(\frac{h\pm i\nu-J-t}{2}\right)$, they yield the poles in $t$-plane corresponding to the double trace operators.
However these $t$-plane poles are canceled by the zeroes from $\Gamma(\delta_{12})\Gamma(\delta_{34}) = \Gamma\left(\frac{\ttau_1+\ttau_2-t}{2}\right) \Gamma\left(\frac{\ttau_3+\ttau_4-t}{2}\right)$ in the denominator.
More explicitly, for example using the results in \cite{SpinningWitten}, 
the the relevant singularities are contained in the dynamical pre-factors $\mathbf{b}({\bf k}_L,{\bf n}_L)$ through $\Gamma$-function of the form:
\be\label{double trace gamma}
\Gamma\left(\frac{\tilde{\tau}_1+\tilde{\tau}_2-(h+i\nu-J)}{2}+N\right)\,,
\ee
where $N$ is some non-negative constant. 
Comparing with the corresponding $\Gamma$-function for the derivation of scalar Mellin amplitude $\Gamma\left(\frac{\Del_1+\Del_2 - (h+i\nu-J)}{2}\right)$,
the additional integer shift here is caused by the additional derivatives and index contractions in the interaction vertices.
The poles in the $\nu$-plane are at $h+i\nu=\tilde{\tau}_1+\tilde{\tau}_2+J +2 N +2m$, and $m$ is also a non-negative integer.
When colliding with the poles in  $\Gamma(\hat{\delta}_{12})$, they yield the
poles in the $t$ plane at $t=\tilde{\tau}_1+\tilde{\tau}_2+2N +2m$.
These poles in the $t$ plane are canceled with zeros which come from {the $\Gamma$-function in the denominator in \eqref{SpinningMellinAmp} $\Gamma(\delta_{12})$ at $t=\tilde{\tau}_1+\tilde{\tau}_2+2m'$, where $m' = 0,1,2,3,\dots$ includes all non-negative integers. 
On the other hand, as similar as the scalar case, when we consider geodesic diagrams, the coefficients do not contain such $\Gamma$-functions including the double trace poles
as in \eqref{double trace gamma}. According to this fact, spinning geodesic diagrams contain only the single trace exchange, and this is consistent with the fact that a geodesic diagram
is proportional to a conformal partial wave, which is associated with the exchange of single type of primary operators. For details about spinning geodesic Witten diagrams, please see \cite{SpinningGWD}, \cite{Castro1} and \cite{Dyer1} and \cite{Nishida1}.}
\paragraph{}
Finally as the scalar case in \eqref{MellinExpansion}, 
the spinning Mellin amplitude \eqref{SpinningMellinAmp} again has Laurent expansion in $t$-channel,
which arises from the same mechanism of the collision of the poles in $\nu$ integration $\nu = \pm i (\Del-h)$
and poles in the $t$ integration of the gamma functions $\Gamma(\hat{\delta}_{12})$ and $\Gamma(\hat{\delta}_{34})$.
The spinning Mellin amplitude has the following expansion:
\bea
&&\int^\infty_{-\infty}\! d\nu ~ b^{({\bf n}_L,{\bf n}_R)}_J(\nu) \cM_{\nu,J}^{({\bf n}_L,{\bf n}_R)}(s,t;a_{ij},b_{ij}) =\sum_{m=0}^\infty \frac{{\bf Q}_{J,m}^{(\bn_L,\bn_R)}(s)}{t-(\Delta-J + 2m)}+ {\rm regular}\,,
\eea
\bea
{\bf Q}_{J,m}^{(\bn_L,\bn_R)}(s)&=& \pi^h g_{\Xi_1\Xi_2\Xi_0}g_{\Xi_3\Xi_4\tilde{\Xi}_0}\left(\prod_{i=1}^4\frac {m_i!}{\Gamma\left(\gamma_{0i}\right)}\right)
\frac{\Gamma(h-\Delta-m)}{\Gamma\left(\frac{\tilde{\tau}_1+\tilde{\tau}_2-\Delta+J}{2}-m\right)
\Gamma\left(\frac{\tilde{\tau}_3+\tilde{\tau}_4-\Delta+J}{2}-m\right)}\nn\\
&&\qquad \times \left[
\mathbf{b}({\bf k}_L,{\bf n}_L)  
\mathbf{b}({\bf k}_R,{\bf n}_R)
\tilde{P}_{\nu,J}^{({\bf n}_L,{\bf n}_R)}(\delta_{ij},a_{ij},b_{ij})\right]_
{\begin{smallmatrix}
\!\!\!\!\!\! \nu=i(h-\Delta)\\
t=\Delta-J+2m
\end{smallmatrix}}.
\eea
As expected, these $t$-channel poles again correspond to the exchange of symmetric traceless primary operator with twist $\ttau=\Del-J$ and its infinite descendants.

\subsection{Simple Examples}
\paragraph{}
In the previous section, we have seen the general expression of the conformal integral with spinning fields.
The general expression is somewhat complicated, in this section we will consider some simple examples of Mellin representations for specific conformal integrals.

\subsubsection*{Example-1: ${\bf n}_L={\bf n}_R={\bf 0}$ }
The first example is the conformal integral of two three point functions which contain only $V_{i,jk}$ and $P_{ij}$.
\bea
\cI^{({\bf 0},{\bf 0})}_{\nu, J}(P_i, Z_i)
=\frac{1}{J!\, (h-1)_J}\int_{\partial}dP_0 
\begin{bmatrix}
~\Delta_1~ &~ \Delta_2 ~&~ h+i\nu~\\
l_1 & l_2 & J\\
0& 0 & 0
\end{bmatrix}
\cdot
\begin{bmatrix}
~\Delta_3~ &~ \Delta_4 ~&~ h-i\nu~\\
l_3 & l_4 & J\\
0& 0 & 0
\end{bmatrix}\,.
\eea
In this case, it is easier to perform the integration, and we obtain the following expression:
\bea\label{example1}
\cI^{({\bf 0},{\bf 0})}_{\nu, J}(P_i, Z_i)&=&
\pi^h \left(\prod_{i=1}^4\frac{l_i!}{\Gamma(\tilde{\gamma}_{0i})}\right)
\sum_{a_{ij},b_{ij}}\int^{i\infty}_{-i \infty} \frac{d\delta_{12}d \delta_{13}}{(2\pi i)^2} \\
&&\qquad \times \tilde{P}^{({\bf 0},{\bf 0})}_{\nu.J}(\delta_{ij},a_{ij},b_{ij})
\frac{\Gamma(\hat{\delta}_{12})
\Gamma(\hat{\delta}_{34})}{\Gamma(\delta_{12})\Gamma(\delta_{34})}
 \prod_{i\neq j}\frac{\tilde{V}_{ij}^{a_{ij}}}{a_{ij}!}\prod_{i<j}
\frac{\tilde{H}_{ij}^{b_{ij}}}{b_{ij}!}\Gamma(\delta_{ij})P_{ij}^{-\delta_{ij}}\,.\nn
\eea
Now $\tilde{\gamma}_{ij}$ are the same as \eqref{def:tilde gamma} for $m_i=l_i$, $m_0^\pm=J$ and $n_{ij}=0$\,.
$\hat{\delta}_{12}$ and $\hat{\delta}_{34}$ are
\bea
\hat{\delta}_{12}=\delta_{12}-\tilde{\gamma}_{12}\,,\qquad \hat{\delta}_{34}=\delta_{34}-\tilde{\gamma}_{34}\,.
\eea
Here the generalized Mack polynomial $\tilde{P}^{({\bf 0},{\bf 0})}_{\nu.J}(\delta_{ij},a_{ij},b_{ij})$ is given as:
\bea
\tilde{P}^{({\bf 0},{\bf 0})}_{\nu.J}(\delta_{ij},a_{ij},b_{ij})&=&\frac{1}{(-2)^{\sum_{i=1}^4l_i}}\sum^{[J/2]}_{r=0}(-1)^r\frac{J!(J+h-1)_{-r}}{r!\, 2^{J}} 
(\hat{\delta}_{12})_{r+b_{12}}(\hat{\delta}_{34})_{r+b_{34}} \\
&\times& \sum_{\sum k_{ij}=J-2r}(-1)^{k_{24}+k_{13}}\prod_{(ij)}\frac{(\delta_{ij})_{k_{ij}+d_{ij}}}{k_{ij}!}
\prod_{i=1}^4\Bigl(\tilde{\gamma}_{0i}-J+r+\sum_j k_{ji}\Bigr)_{J-r-\sum_{j}k_{ji}}
\,.\nn
\eea
This polynomial has almost the same form as the scalar case \eqref{Def:tildeP}, 
except for some additional Pochhammer symbols and the over all factors.
The definition of $d_{ij}$ is the same as in \eqref{shift:tilde delta}.

\subsubsection*{Example-2: ${\bf n}_L={\bf n}_R={\bf n}_{(J)}=(J,0,0)$ }
\paragraph{}
Next we consider the case with ${\bf n}_L={\bf n}_R={\bf n}_{(J)}=(J,0,0)$\,:
\bea
\cI^{({\bf n}_{(J)},{\bf n}_{(J)})}_{\nu, J}(P_i, Z_i)
\equiv\frac{1}{J!\, (h-1)_J}\int_{\partial}dP_0 
\begin{bmatrix}
~\Delta_1~ &~ \Delta_2 ~&~ h+i\nu~\\
l_1 & l_2 & J\\
J & 0 & 0
\end{bmatrix}
\cdot
\begin{bmatrix}
~\Delta_3~ &~ \Delta_4 ~&~ h-i\nu~\\
l_3 & l_4 & J\\
J & 0 & 0
\end{bmatrix}\,.
\eea
Here we suppose that $l_2\geq J$ and $l_4\geq J$ to keep $m_i$ non-negative integers. 
The differential operator $\cD_{Z_0}$ and the polarization vector $Z_0$ are contained only in $H_{02}$ and $H_{04}$ respectively,
and the contraction between $\cD_{Z_0}$ and $Z_0$ are evaluated only through $H_{02}$ and $H_{04}$ instead of $V_{012}$ and $V_{0,34}$ 
as in the scalar case  \eqref{Z-contraction}.
Now we have to evaluate the following combination:
\bea
&&\frac{1}{J!(h-1)_J}\left(2P_0\cdot C_2 \cdot \cD_{Z_0}\right)^J \left(2P_0\cdot C_4 \cdot Z_0\right)^J\,.
\eea
This type of combination gives the Gegenbauer polynomial in general as in \eqref{Z-contraction}. 
In this case, however, because $(-4P_0 \cdot C_i\cdot C_i \cdot P_0)=0$, 
the terms of the polynomial are reduced, and the combination is equivalent to $(-4P_0 \cdot C_2\cdot C_4 \cdot P_0)^J$.
After the similar calculation, we can obtain the following generalized Mack polynomial:
\bea\label{mack ex2}
&&\tilde{P}^{({\bf n}_{(J)},{\bf n}_{(J)})}_{\nu.J}(\delta_{ij},a_{ij},b_{ij})= 
\frac{J!}{2^J}\sum_{\sum k=J}(-1)^{k_{24}+k_{\bar{2}\bar{4}}}\frac{1}{(-2)^{\sum_{i}l_i-2J}}
\\
&&\quad \times 
(m_2+1)_{\bar{\kappa}_2}
(m_4+1)_{\bar{\kappa}_4}
(\tilde{\gamma}_{02}-\kappa_2)_{\kappa_2}
(\tilde{\gamma}_{04}-\kappa_4)_{\kappa_4}
(\hat{\delta}_{12})_{b_{12}}
(\hat{\delta}_{34})_{b_{34}}
\nn\\
&&\quad \times 
(\delta_{13})_{d_{13}} 
(\delta_{14})_{d_{14}} 
(\delta_{23})_{d_{23}} 
\frac{(\delta_{24})_{d_{24}+k_{24}}}{k_{24}!}
\frac{(a_{24}-k_{\bar{2}4}+1)_{k_{\bar{2}4}}}{k_{\bar{2}4}!}
\frac{(a_{42}-k_{\bar{4}2}+1)_{k_{\bar{4}2}}}{k_{\bar{4}2}!}
\frac{(b_{24}-k_{\bar{2}\bar{4}}+1)_{k_{\bar{2}\bar{4}}}}{k_{\bar{2}\bar{4}}!}\,.\nn
\eea
Note that there is no $r$-summation because of the reduction of the Gegenbauer polynomial,
and we have only the summation in $k$.
In this example, only $\beta^{\{\bar{2}2\bar{4}4\}}(=J)$ is non-zero, and 
these $k_{\alpha\beta}$ in \eqref{mack ex2} correspond to $k_{\alpha\beta}^{\{\bar{2}2\bar{4}4\}}$ in the general case.
$\kappa_i$ and $\bar{\kappa}_i$ are given as 
\bea
&&\kappa_2= J -k_{24}-k_{2\bar{4}}\,,\quad \kappa_4= J -k_{24}-k_{\bar{2}4}\,,\nn\\
&&\bar{\kappa}_2= J -k_{\bar{2}4}-k_{\bar{2}\bar{4}}\,,\quad \bar{\kappa}_4= J -k_{2\bar{4}}-k_{\bar{2}\bar{4}}\,, 
\eea 
and the others are zero. The definition of $d_{ij}$ is the same as before.
\paragraph{}

Here we end with a couple of  short comments about the possible future directions.
In this paper we considered only the Mellin representation of the spinning four point functions, it would be interesting to also consider the higher point functions involving symmetric traceless tensor fields.
Although it is still difficult to obtain the explicit form, we can count the number of independent variables through 
the generalized Symanzik formula which is useful even for $n$-point functions.
Counting the number of discrete Mellin variables arising in this general case, there are altogether $\frac{3n(n-1)}{2}$ of $\{a_{ij},b_{ij}\}$\footnote{
Here we assumed that $n<d$, where $d$ is the number of the Euclidean spacetime dimensions.},
however among them there are also $n$ constraints, 
the numbers of independent discrete Mellin variables is $\frac{n(3n-5)}{2}$.
Again, this number matches with the counting of flat space independent elements of tensor structures.
It would also be interesting to generalize our analysis to more general representations, while here we consider only symmetric traceless operators.
There are already some works related to such a direction \cite{FermionicMellin,MixedTensor1,GeneralBlock2}. 

\acknowledgments
{The authors would like to thank Massimo Taronna and Charlotte Sleight for for helpful comments on our draft, and we are also grateful to Massimo Taronna for useful discussions.}
This work was supported in part by National Science Council through the grant  104-2112-M-002 -004 -MY, Center for Theoretical Sciences at National Taiwan University. 
HYC would like to thank Organizers of 2017 KEK theory workshop for the opportunity to present part of this work.
The work of HK is supported by the Japan Society for the Promotion of Science (JSPS) and by the Supporting Program for Interaction-based Initiative Team Studies (SPIRITS) from Kyoto University.

\appendix
\section*{Appendix}


\section{Symanzik Star-Formulae}\label{App:Symanzik}
\paragraph{}
In this section, {we summarize the Symanzik star-formula \cite{Symanzik}} and its generalization to the spinning case, these are used to derive the corresponding Mellin amplitudes from the boundary or bulk integration.
\subsection{Normal Symanzik Star-Formula}\label{App:Normal-Symanzik}
\paragraph{}
Here we give a short review of some useful integration formulae using the Schwinger parametrization to calculate the boundary and bulk integrals.  
Firstly we consider the following $n$-point boundary integration, after using Schwinger parametrization and performing the $P_0$ integration, it is evaluated as {(For the detail, please see \cite{MellinAmp1})}:
\bea\label{SF1}
\int_{\partial} dP_0~ \prod_{i=1}^n \frac{1}{(-2 P_i\cdot P_{0})^{\delta_i}}=2\pi^h\left(\prod_{i=1}^n \frac{1}{\Gamma(\delta_i)} \int_0^\infty \frac{d t_i}{t_i}~
t_i^{\delta_i} \right)  e^{-\sum_{i<j} t_it_jP_{ij}}\,.
\eea
Here we have assumed that $\sum_{i=1}^n \delta_i=d$ in this boundary integral.
Similarly, for the bulk integration, we can obtain: 
\bea\label{SF2}
\!\!\!\!\!\! \int_{\rm AdS} dX~ \prod_{i=1}^n\frac{1}{(-2P_i\cdot X)^{\delta_i}}=\pi^h \Gamma\left(\frac{\sum_{i=1}^n \delta_i -d}{2 }\right) \left(\prod_{i=1}^n\frac{1}{\Gamma(\delta_i)}
\int_0^\infty  \frac{d t_i}{t_i}~
t_i^{\delta_i}\right) e^{-\sum_{i<j} t_it_j P_{ij}}\,,
\eea
notice we do not impose the condition $\sum_{i=1}^n \delta_i =d$ in the bulk integration despite the notation.
Next we consider the remaining integration of the Schwinger parameters $\{t_i\}$. 
Changing the integration variables appropriately and using the following formula:
\bea\label{SF3}
e^{-t_it_j P_{ij}} = \int_{-i\infty}^{i\infty}\frac{d \delta_{ij}}{2 \pi i} ~\Gamma(\delta_{ij})(t_it_jP_{ij})^{-\delta_{ij}}\,,
\eea
we can rewrite the integration, using the Mellin coordinates $\{\delta_{ij}\}$:
\bea\label{SF4}
\left(\prod_{i=1}^n\int_0^\infty \frac{d t_i}{t_i} ~t_i^{\delta_i}\right)  e^{-\sum_{i<j} t_it_jP_{ij}}=\frac{1}{2}\int_{-i\infty}^{i\infty}[d\delta]_{\frac{n(n-3)}{2}}\prod_{i<j}\Gamma(\delta_{ij})P_{ij}^{-\delta_{ij}}\,,
\eea
where the integration measure is defined as:
\bea\label{SF5}
[d\delta]_{\frac{n(n-3)}{2}} = \frac{\prod_{[ij]}d\delta_{ij}}{(2\pi i)^{\frac{n(n-3)}{2}}}\,,
\eea
and $[ij]$ label the $\frac{n(n-3)}{2}$ combinations of $i$ and $j$.
{Note here the number of the independent $\{\delta_{ij}\}$ is $\frac{n(n-3)}{2}$ due to the following relation:}
\bea\label{SF6}
\sum_{j(\neq i)}\delta_{ij}=\delta_i\,.
\eea
As a summary, the boundary and bulk integration have the following Mellin representation:
\bea\label{SF7}
\int_{\partial} dP_0~ \prod_{i=1}^n \frac{1}{(-2P_i\cdot P_0)^{\delta_i}}&=&\pi^h\left(\prod_{i=1}^n \frac{1}{\Gamma(\delta_i)}\right) \int_{-i\infty}^{i\infty}[d\delta]_{\frac{n(n-3)}{2}}\prod_{i<j}\Gamma(\delta_{ij})P_{ij}^{-\delta_{ij}}\,,\\
\int_{\rm AdS} dX~ \prod_{i=1}^n \frac{1}{(-2P_i\cdot X)^{\delta_i}} &=&\frac{\pi^h}{2} \Gamma\left(\frac{\sum_{i=1}^n \delta_i -d}{2 }\right) \left(\prod_{i=1}^n\frac{1}{\Gamma(\delta_i)}\right)
\int_{-i\infty}^{i\infty}[d\delta]_{\frac{n(n-3)}{2}}\prod_{i<j}\Gamma(\delta_{ij})P_{ij}^{-\delta_{ij}}\,.\nn\\
\eea

\subsection{Generalized Symanzik Star-Formula}\label{App:Generalized Symanzik}
\paragraph{}
Here we will generalize the Symanzik star-formula \cite{Symanzik} discussed in the previous section to facilitate the computation of spinning Mellin amplitudes\footnote{
{
{This type of generalization was also considered in \cite{generalized Symanzik} or \cite{LoopMellin}.}
}
}.
Firstly, we consider the following boundary integration including light-like vectors $Y_i$ satisfying $Y_i\cdot Y_i = P_i\cdot Y_i =0$:
\bea\label{GSF-bdy-1}
\int_{\partial} dP_0~ \prod^{n}_{i=1} \frac{(Y_i\cdot P_0)^{\xi_i}}{(-2 P_i \cdot P_0)^{\delta_i}}\,,
\eea
where $\xi_i$ are positive integers and we assume $\sum_{i=1}^n(\delta_i-\xi_i)=d$\,.
To calculate this integral, we use the Schwinger parametrization for each $(P_{i0})^{-\delta_i}$, and for $(Y_i\cdot P_0)^{\xi_i}$ we use the following relation:
\bea\label{GSF-bdy-2}
(Y_i\cdot X)^{\xi_i} = \xi_i ! \oint \frac{d \zeta_i}{2 \pi i \, \zeta_i} ~\zeta_i^{-\xi_i} e^{\zeta_i Y_i\cdot X }\,,
\eea
where the contour of integral is a small circle around the origin. 
After performing $P_0$ integral, we obtain:
\bea\label{GSF-bdy-3}
\int_{\partial} dP_0~ \prod^{n}_{i=1} \frac{(Y_i\cdot P_0)^{\xi_i}}{(-2 P_i \cdot P_0)^{\delta_i}}=2\pi^h  \left(\prod^{n}_{i=1}\frac{\xi_i!}{\Gamma(\delta_i)}\int_0^{\infty}\frac{dt_i}{t_i}\oint \frac{d\zeta_i}{2\pi i \, \zeta_i}~t_i^{\delta_i}\, \zeta_i^{-\xi_i}\right)
 e^{-|Q|^2}\,,
\eea
where the combined vector $Q^A$ is given by
\bea\label{GSF-bdy-4}
Q^A \equiv \sum_{i=1}^n t_i~ P_i^A+\frac{1}{2}\sum_{i=1}^n \zeta_i ~Y_i^A\,.
\eea
Next we consider the bulk integration whose integrand is the same as \eqref{GSF-bdy-1} except with $X^A$ exchanged with $P_0^A$:
\bea\label{GSF-bulk-1}
\int_{\text{AdS}} dX~ \prod^{n}_{i=1} \frac{(Y_i\cdot X)^{\xi_i}}{(-2 P_i \cdot X)^{\delta_i}}\,.
\eea
Using the Schwinger parametrization and \eqref{GSF-bdy-2}, the $X$ integration is evaluated as:
\be\label{GSF-bulk-2}
\int_{\text{AdS}} dX~ \prod^{n}_{i=1} \frac{(Y_i\cdot X)^{\xi_i}}{(-2 P_i \cdot X)^{\delta_i}}
=\pi^h  \Gamma\left(\frac{\sum_{i=1}^n (\delta_i-\xi_i)-d}{2}\right)
 \left(\prod^{n}_{i=1}\frac{\xi_i!}{\Gamma(\delta_i)}\int_0^{\infty}\frac{dt_i}{t_i}\oint \frac{d\zeta_i}{2\pi i \, \zeta_i}~t_i^{\delta_i}\, \zeta_i^{-\xi_i}\right)
 e^{-|Q|^2}\,.
\ee
where the vector $Q^A$ is the same as \eqref{GSF-bdy-4}\,.
\paragraph{}
Similar to the previous section, the integration over the Schwinger parameters in \eqref{GSF-bdy-3} and \eqref{GSF-bulk-2} can be rewritten by using the Mellin integration:
\bea
\!\!\!\!\!\! \tilde{\cI}(P_i,Y_i)= \left(\prod^{n}_{i=1}\int_0^{\infty}\frac{dt_i}{t_i}\oint \frac{d\zeta_i}{2\pi i \, \zeta_i}~t_i^{\delta_i}\, \zeta_i^{-\xi_i}\right)
 e^{-\sum_{i<j} (t_i t_j P_{ij}-\frac{1}{2}\zeta_i \zeta_j ( Y_i\cdot Y_j)) +\sum_{i \neq j} t_i \zeta_j (Y_i \cdot P_j)}\,.
\eea
For the product $( Y_i\cdot Y_j)$ and $(Y_i \cdot P_j)$, we insert the expansion of exponential function:
\bea
e^{ t_i \zeta_j (Y_i \cdot P_j)}&=&\sum_{a_{ij}=0}^\infty \frac{1}{a_{ij}!} (t_i \, \zeta_j)^{a_{ij}} (Y_i\cdot P_j)^{a_{ij}}\,,\nn\\
e^{\frac{1}{2} \zeta_i \zeta_j (Y_i \cdot Y_j)}&=&\sum_{b_{ij}=0}^\infty \frac{1}{b_{ij}!} (\zeta_i \, \zeta_j)^{b_{ij}} \left(\frac{1}{2}Y_i\cdot Y_j\right)^{b_{ij}}\,.
\eea
Notice that $\{ b_{ij} \}$ are symmetric: $b_{ij}=b_{ji}$, but $\{a_{ij}\}$ is not.
Then $\tilde{\cI}$ becomes:
\bea
\tilde{\cI}(P_i,Y_i)&=& \sum_{a_{ij}=0}^\infty\sum_{b_{ij}=0}^\infty\left(\prod^{n}_{i=1}\int_0^{\infty}\frac{dt_i}{t_i}\oint \frac{d\zeta_i}{2\pi i \, \zeta_i}
~t_i^{\delta_i+\sum_{j(\neq i)}a_{ij}}\, \zeta_i^{-\xi_i +\sum_{j(\neq i)}(b_{ij}+a_{ji})}\right)\nn\\
&& \prod_{i\neq j} \frac{(Y_i\cdot P_j)^{a_{ij}}}{a_{ij}!} \prod_{i< j} \frac{\left(\frac{1}{2}Y_i\cdot Y_j\right)^{b_{ij}}}{b_{ij}!} e^{-\sum_{i<j}t_i t_j P_{ij}}\,.
\eea
Now $\xi_i$ integral is easily evaluated and this integral is zero unless 
\be
\sum_{j(\neq i)}(b_{ij}+a_{ij})=\xi_i.
\ee
Due to this constraint, the summation over $a_{ij}$ and $b_{ij}$ is restricted into a finite region.
For the $t_i$ integral, we can use the Symanzik star-formula \eqref{SF4}, and the integral becomes:
\bea
\tilde{\cI}(P_i,Y_i)&=& \frac{1}{2}\sum_{a_{ij}, b_{ij}}\int_{-i\infty}^{i\infty}[d\delta]_{\frac{n(n-3)}{2}}
 \prod_{i\neq j} \frac{(Y_i\cdot P_j)^{a_{ij}}}{a_{ij}!}
\prod_{i<j} \frac{\left(\frac{1}{2}Y_i\cdot Y_j\right)^{b_{ij}}}{b_{ij}!}\, \Gamma(\delta_{ij})P_{ij}^{-\delta_{ij}}\,,
\eea
where the Mellin variables $\delta_{ij}$\,, $a_{ij}$ and $b_{ij}$ satisfy the following constraints:
\bea
\sum_{j (\neq i)} (\delta_{ij}-a_{ji})=\delta_i\,,\qquad \sum_{j (\neq i)} (b_{ij}+a_{ij})=\xi_i\,.
\eea
Note that the first condition ensures that the total power of $P_i$ is $-\delta_i$ and the second {ensures} that the total power of $Y_i$ is integer $\xi_i$\,.

\section{The Details of Calculation}
\subsection{Conformal Integral for spinning cases}\label{App:CI Spinning}
\paragraph{}
Here we will see the details of the calculation of general conformal integral in \eqref{general integral 0}.
Using the definition of the box tensor basis and the vectors $\{\bar{v}_i\}$, we can rewrite the kinematical integral as:
\bea\label{general-box}
\cI^{({\bf n}_L,{\bf n}_R)}_{\nu, J}(P_i, Z_i)
&=&\frac{H_{12}^{n_{12}}H_{34}^{n_{34}}}{P_{12}^{\tilde{\gamma}_{12}}P_{34}^{\tilde{\gamma}_{34}}}
\int dP_0 \left(\prod_{i=1}^4\frac{(\bar{v}_i\cdot P_0)^{m_i}}{P_{0i}^{\tilde{\gamma}_{0i}}}\right)
\mathcal{V}({\bf m}_L, {\bf m}_R)\,.
\eea
Now $\tilde{\gamma}_{ij}$ are shifted dimensions as below:
\bea\label{def:tilde gamma}
\tilde{\gamma}_{12}&=&\frac{\tau_1+\tau_2-\tau^+_0}{2}+m^+_0-m_1-m_2\,,\quad 
\tilde{\gamma}_{34}=\frac{\tau_3+\tau_4-\tau^-_0}{2}+m^-_0-m_3-m_4\,,\no\\
\tilde{\gamma}_{01}&=&\frac{\tau_1+\tau^+_0-\tau_2}{2}+m_2\,,\quad 
\tilde{\gamma}_{02}=\frac{\tau_2+\tau^+_0-\tau_1}{2}+m_1\,,\no\\
\tilde{\gamma}_{03}&=&\frac{\tau_3+\tau^-_0-\tau_4}{2}+m_4\,,\quad 
\tilde{\gamma}_{04}=\frac{\tau_4+\tau^-_0-\tau_3}{2}+m_3\,,
\eea
where $\tau_i = \Delta_i+l_i$\,, $\tau^{\pm}_0=h\pm i\nu+J$. 
The vectors $\{\bar{v}_i\}$ are combinations of a polarization vector and coordinates which are defined in \eqref{def:bar v}.
$\mathcal{V}({\bf m}_L, {\bf m}_R)$ is introduced as the contraction part:
\bea
\mathcal{V}({\bf m}_L, {\bf m}_R)
&\equiv& \frac{1}{J!\, (h-1)_J} (2P_0\cdot \mathcal{P}_{12}\cdot \mathcal{D}_{Z_0})^{m^+_0}(2P_0\cdot C_1 \cdot \mathcal{D}_Z)^{n_{01}}(2P_0\cdot C_2 \cdot \mathcal{D}_Z)^{n_{20}}\no\\
&&\qquad\qquad\times
(2P_0\cdot \mathcal{P}_{34}\cdot Z_0)^{m^-_0}(2P_0\cdot C_3 \cdot Z_0)^{n_{03}}(2P_0\cdot C_4 \cdot Z_0)^{n_{40}}\,,
\eea
where ${\bf m}_L=(m^+_0,n_{01},n_{02})$ and ${\bf m}_R=(m^-_0,n_{03},n_{04})$\,.
To calculate the contraction between $\cD_{Z_0}$ and $Z_0$, we introduce the  two following rank-two anti-symmetric tensors:
\be
\mathcal{W}_{12}^{AB}({\bf t}_L)\equiv (t^+_0 \mathcal{P}_{12}+t_1 C_{1}+t_2 C_{2})^{AB},\quad
\mathcal{W}_{34}^{AB}({\bf t}_R)\equiv (t^-_0 \mathcal{P}_{34}+t_3 C_{3}+t_4 C_{4})^{AB}\,,
\ee
where two sets of triplets of real parameters ${\bf t}_L=(t^+_0,t_{1},t_2)$ and ${\bf t}_R=(t^-_0,t_3,t_4)$.
Using $\mathcal{W}_{12}$ and $\mathcal{W}_{34}$, we can rewrite $\mathcal{V}(\bm_L, \bm_R)$ as:
\bea
\mathcal{V}({\bf m}_L, {\bf m}_R)
&=&\frac{1}{J!\, (h-1)_J} \partial_{{\bf t}_L , {\bf t}_R}(2P_0\cdot \mathcal{W}_{12}\cdot \mathcal{D}_Z)^J(2P_0\cdot \mathcal{W}_{34}\cdot Z_0)^J\,.
\eea
Here the $t$-differential operator $\partial_{{\bf t}_L , {\bf t}_R}$ is defined as:
\bea
\partial_{{\bf t}_L , {\bf t}_R}(...)
\equiv  \frac{1}{(J!)^2}
 \partial_{t^+_0}^{m^+_0}\partial_{t_1}^{n_{01}} \partial_{t_2}^{n_{02}}\partial_{t^-_0}^{m^-_{0}}
 \partial_{t_3}^{n_{03}}\partial_{t_4}^{n_{04}}\left. (...)\right|_{{\bf t}_L={\bf 0} , {\bf t}_R={\bf 0}}.
\eea
Now we can easily evaluate the contraction between $\cD_{Z_0}$ and $Z_0$ and it gives Gegenbauer polynomial:
\bea\label{second-2}
&&\cV({\bf m}_L, {\bf m}_R)\nn\\
&&=\partial_{{\bf t}_L , {\bf t}_R}
\tilde{\sum_r}(-4P_0\cdot \mathcal{W}_{12}\cdot\mathcal{W}_{34}\cdot P_0)^{J-2r}
(-4P_0\cdot \mathcal{W}_{12}\cdot \mathcal{W}_{12}\cdot P_0)^{r}(-4P_0\cdot \mathcal{W}_{34}\cdot \mathcal{W}_{34}\cdot P_0)^{r}\nn\\
&&=\tilde{\sum_r}\sum_{\beta^{\{A,L,R\}}}\frac{m^+_0!n_{01}!n_{02}!m_0^-! n_{03}!n_{04}!}{(J!)^2}\frac{(J-2r)!}{\prod_{\{A\}}\beta^{\{A\}}!}\mathcal{V}^{(1234)}
 \frac{r!}{\prod_{\{L\}}\beta^{\{L\}}!}\mathcal{V}^{(1212)}\frac{r!}{\prod_{\{R\}}\beta^{\{R\}}!}\mathcal{V}^{(3434)}\,,\nn\\
\eea
The summation $\tilde{\sum}_{r}$ is the same as defined in \eqref{ShortSummation}.
In the second line, we expanded the factors $(-4P_0\cdot \mathcal{W}_{12}\cdot\mathcal{W}_{34}\cdot P_0)$, $(-4P_0\cdot \mathcal{W}_{12}\cdot\mathcal{W}_{12}\cdot P_0)$ and $(-4P_0\cdot \mathcal{W}_{34}\cdot\mathcal{W}_{34}\cdot P_0)$\footnote{We will label the powers of the various terms obtained from the polynomial expansion by the indices $\beta^{\{...\}}$. see \eqref{v1234} and \eqref{def:A L R}.}, and applied the differential operator $\partial_{{\bf t}_L , {\bf t}_R}$\,.
Here the factors $\{\mathcal{V}^{(....)}\}$ are given as:
\bea\label{v1234}
\mathcal{V}^{(1234)}&=&
(-4 P_0\cdot \mathcal{P}_{12}\cdot \mathcal{P}_{34}\cdot P_0)^{\beta^{\{1234\}}}
(-4 P_0\cdot C_1 \cdot \mathcal{P}_{34}\cdot P_0)^{\beta^{\{1\bar{1}34\}}}
(-4 P_0\cdot  C_{2}\cdot \mathcal{P}_{34}\cdot P_0)^{\beta^{\{\bar{2}234\}}}\no\\
&\times &
(-4P_0\cdot \mathcal{P}_{12} \cdot C_3\cdot P_0)^{\beta^{\{123\bar{3}\}}}
(-4P_0\cdot \mathcal{P}_{12} \cdot C_4\cdot P_0)^{\beta^{\{12\bar{4}4\}}}
(-4 P_0\cdot  C_{1}\cdot C_3\cdot P_0)^{\beta^{\{1\bar{1}3\bar{3}\}}}
\no\\
&\times &
(-4 P_0\cdot  C_{2}\cdot C_3\cdot P_0)^{\beta^{\{\bar{2}23\bar{3}\}}}
(-4 P_0\cdot  C_{1}\cdot C_4\cdot P_0)^{\beta^{\{1\bar{1}\bar{4}4\}}}
(-4 P_0\cdot  C_{2}\cdot C_4\cdot P_0)^{\beta^{\{\bar{2}2\bar{4}4\}}}\,,
\no\\[8pt]
\mathcal{V}^{(1212)}&=&(-4P_0\cdot \mathcal{P}_{12} \cdot \mathcal{P}_{12}\cdot P_0)^{\beta^{\{1212\}}}
(-8 P_0\cdot  \mathcal{P}_{12} \cdot C_1\cdot P_0)^{\beta^{\{121\bar{1}\}}}
\no\\
& \times &
(-8 P_0\cdot  \mathcal{P}_{12}  \cdot C_2\cdot P_0)^{\beta^{\{12\bar{2}2\}}}
(-8 P_0\cdot  C_1 \cdot C_2\cdot P_0)^{\beta^{\{1\bar{1}\bar{2}2\}}}\,,\no\\[8pt]
\mathcal{V}^{(3434)}&=&(-4P_0\cdot \mathcal{P}_{34} \cdot \mathcal{P}_{34}\cdot P_0)^{\beta^{\{3434\}}}
(-8 P_0\cdot  \mathcal{P}_{34} \cdot C_3\cdot P_0)^{\beta^{\{343\bar{3}\}}}\no\\
& \times &
(-8 P_0\cdot  \mathcal{P}_{34}  \cdot C_4\cdot P_0)^{\beta^{\{34\bar{4}4\}}}
(-8 P_0\cdot  C_3 \cdot C_4\cdot P_0)^{\beta^{\{3\bar{3}\bar{4}4\}}}\,.
\eea
In \eqref{second-2}, we have used the abridged notations:
\bea\label{def:A L R}
\{A\} &=& \Bigl\{\,\{1234\},~\{1\bar{1}34\},~\{\bar{2}234\},~\{123\bar{3}\},~\{12\bar{4}4\},~\{1\bar{1}3\bar{3}\},~\{\bar{2}234\},~\{1\bar{1}\bar{4}4\},~\{\bar{2}2\bar{4}4\}\Big\}\,,\nn\\
\{L\} &=& \Bigl\{\,\{1212\},~\{121\bar{1}\},~\{12\bar{2}2\},~\{1\bar{1}\bar{2}2\} \, \Bigr\}\,,\nn\\
\{R\} &=&  \Bigl\{\,\{3434\},~\{343\bar{3}\},~\{34\bar{4}4\},~\{3\bar{3}\bar{4}4\} \, \Bigr\}\,.
\eea 
to denote the sets of indices arising from the expansions of $\cV^{(1234)}$, $\cV^{(1212)}$ and $\cV^{(3434)}$ respectively. 
Notice that we have used combined indices $ij$  to associate with the anti-symmetric tensor ${\mathcal P}_{ij}$, 
however the ordered combined indices $i\bar{i}$ and $\bar{i} i$ indicate that we can obtain $C_i$ from $\cP_{ij}$ or from $\cP_{ji}$ by replacing $P_j$ with $\bar{v_i}$.
Note that seventeen possible {non-negative integers} $\beta^{\{...\}}$ need to satisfy the following nine constraints and the summations of $\beta^{\{...\}}$ in the last line of \eqref{second-2} 
are taken within the values where the constraints are satisfied\footnote{
For example, the first constraint is written explicitly as follows:
\bea
\beta^{\{1234\}}+\beta^{\{123\bar{3}\}}+\beta^{\{12\bar{4}4\}}+2\beta^{\{1212\}}+ \beta^{\{121\bar{1}\}}+\beta^{\{12\bar{2}2\}}= m_0^+\,.
\eea
Note here that there is factor 2 in front of $\beta^{\{1212\}}$ because it contains two set of $12$\,, and $\beta^{\{1\bar{1}\bar{2}2\}}$ is not included in the second summation
because it does not contain $\cP_{12}$\,. 
}:
\bea
\sum_{\{..12..\}\in \{A\}}\beta^{\{A\}}+\sum_{\{..12..\}\in \{L\}}\beta^{\{L\}}&=&m^+_0\,,\quad
\sum_{\{..1\bar{1}..\}\in \{A\}}\beta^{\{A\}}+\sum_{\{..1\bar{1}..\}\in \{L\}}\beta^{\{L\}}=n_{01}\,,\nn\\
\sum_{\{..\bar{2}2..\}\in \{A\}}\beta^{\{A\}}+\sum_{\{..\bar{2}2..\}\in \{L\}}\beta^{\{L\}}&=&n_{02}\,,\quad
\sum_{\{..34..\}\in \{A\}}\beta^{\{A\}}+\sum_{\{..34..\}\in \{R\}}\beta^{\{R\}}=m^-_{0}\,,\nn\\
\sum_{\{..3\bar{3}..\}\in \{A\}}\beta^{\{A\}}+\sum_{\{..3\bar{3}..\}\in \{R\}}\beta^{\{R\}}&=&n_{03}\,,\quad
\sum_{\{..\bar{4}4..\}\in \{A\}}\beta^{\{A\}}+\sum_{\{..\bar{4}4..\}\in \{R\}}\beta^{\{R\}}=n_{04}\,,\nn\\
\sum_{\{A\}}\beta^{\{A\}}=J-2r\,,\quad \sum_{\{L\}} \beta^{\{L\}}&=&r\,,\quad \sum_{\{R\}}\beta^{\{R\}}=r\,.
\eea
The first six constraints come from the differentiations with respect to parameters ${\bf t}_L$ and ${\bf t}_R$\,, and the last three constraints come from the expansion of polynomials.
Next we have to expand each factor in \eqref{v1234}\,. For example, one of the factors, $(-4 P_0\cdot \mathcal{P}_{12}\cdot \mathcal{P}_{34}\cdot P_0)$\,, is expanded as:
\bea \label{expansion 1234}
&&(-4 P_0\cdot \mathcal{P}_{12}\cdot \mathcal{P}_{34}\cdot P_0)^{\beta^{\{1234\}}}\nn\\
&&=\Bigl( \frac{1}{2} \left( P_{14}P_{02}P_{03}-P_{13}P_{02}P_{04}-P_{24}P_{01}P_{03}+P_{23}P_{01}P_{04}\right)\Bigr)^{\beta^{\{1234\}}}\nn\\
&&=\frac{\beta^{\{1234\}}!}{2^{\beta^{\{1234\}}}} 
\sum_{\sum k^{\{1234\}}_{ij}=\beta^{\{1234\}}}(-1)^{k^{\{1234\}}_{(1)(3)}+k^{\{1234\}}_{(2)(4)}}\prod_{(ij)}\frac{P_{ij}^{k^{\{1234\}}_{ij}}}{k^{\{1234\}}_{ij}!} \prod_{i} P_{0i}^{\beta^{\{1234\}}-{\sum_j k^{\{1234\}}_{ij}}}\,.
\eea
Here we introduce new four integers $k^{\{1234\}}_{ij}$ labeling the partition of $\beta^{\{1234\}}$\,. 
The indices of ${(1)(3)}$ in the third line means the first and third indexes in the bracket are substituted, for example, $k^{\{ijkl\}}_{(1)(3)}=k^{\{ijkl\}}_{ik}$,
and $(ij)$ runs over only $(13)$, $(14)$, $(23)$ and $(24)$\,.
\paragraph{}
The expansions of the other factors can be obtained by replacing $P_i$ with $\bar{v}_i$ appropriately according to the bracket index.
For example, the expansion of $(-4 P_0\cdot \mathcal{P}_{12}\cdot C_3\cdot P_0)$ is obtained by replacing $P_4$ with $\bar{v}_3$:
\bea
&&(-4 P_0\cdot \mathcal{P}_{12}\cdot C_3\cdot P_0)^{\beta^{\{123\bar{3}\}}}\nn\\
&&=\Bigl( \frac{1}{2}\left((-2\bar{v}_3\cdot P_1)P_{02}P_{03}-P_{13}P_{02}(-2\bar{v}_3\cdot P_0)- (-2\bar{v}_3 \cdot P_2)P_{01} P_{03}+P_{23}P_{01} (-2\bar{v}_3\cdot P_0)\right)\Bigr)^{\beta^{\{123\bar{3}\}}}\nn\\
&&=\frac{\beta^{\{123\bar{3}\}}!}{2^{\beta^{\{123\bar{3}\}}}} 
\sum_{\sum k^{\{123\bar{3}\}}_{\alpha\beta}=\beta^{\{123\bar{3}\}}}(-1)^{k^{\{123\bar{3}\}}_{(1)(3)}+k^{\{123\bar{3}\}}_{(2)(4)}}
\prod_{(\bar{i}j)\in \{123\bar{3}\}}\frac{(-2 \bar{v}_{i}\cdot P_j)^{k^{\{123\bar{3}\}}_{\bar{i}j}}}{k^{\{123\bar{3}\}}_{\bar{i}j}!}\nn\\
&&\quad\times\!\!\prod_{(ij)\in \{123\bar{3}\} }\frac{P_{ij}^{k^{\{123\bar{3}\}}_{ij}}}{k^{\{123\bar{3}\}}_{ij}!} 
\prod_{i\in \{123\bar{3}\}} P_{0i}^{\beta^{\{123\bar{3}\}}-\sum_\alpha k^{\{123\bar{3}\}}_{i\alpha}}
\prod_{\bar{i}\in \{123\bar{3}\}} (-2\bar{v}_i\cdot P_0)^{\beta^{\{123\bar{3}\}}-\sum_\alpha k^{\{123\bar{3}\}}_{\bar{i}\alpha}}
\,.
\eea
Similarly as in the previous example, we introduced four non-negative integers $k^{\{123\bar{3}\}}_{\alpha\beta}$ to denote the four-folds partition of $\beta^{\{123\bar{3}\}}$.
The indices $\alpha$ and $\beta$ can be $i$ and $\bar{i}$ which take values in $\{123\bar{3}\}$, such that in this case $i =1,2,3$ in the third product and $\bar{i} =\bar{3}$ in the fourth product in the summation.
And following these assignments, we have $k^{\{123\bar{3}\}}_{1\bar{3}}$ and $k^{\{123\bar{3}\}}_{2\bar{3}}$ from the first product and 
$k^{\{123\bar{3}\}}_{13}$ and $k^{\{123\bar{3}\}}_{23}$ from the second product in the summation. 
Comparing this case with \eqref{expansion 1234},
index $4$ is replaced with $\bar{3}$ because this polynomial expansion is obtained by replacing $P_4$ with $\bar{v}_3$.
\paragraph{}
The next example is the expansion of the factor such as $(-8 P_0\cdot  C_1 \cdot C_2\cdot P_0)^{\beta^{\{1\bar{1}\bar{2}2\}}}$ {in the expansion of $\cV^{(1212)}$}:
\bea
(-8 P_0\cdot  C_1 \cdot C_2\cdot P_0)^{\beta^{\{1\bar{1}\bar{2}2\}}}
&=&\left(P_{12}(-2 \bar{v}_1\cdot P_0)(-2 \bar{v}_2\cdot P_0) + (-2 \bar{v}_1\cdot \bar{v}_2)P_{01}P_{02}\right)^{\beta^{\{1\bar{1}\bar{2}2\}}}\\
&=&\sum_{k^{\{1\bar{1}\bar{2}2\}}+\bar{k}^{\{1\bar{1}\bar{2}2\}}=\beta^{\{1\bar{1}\bar{2}2\}}}\frac{\beta^{\{1\bar{1}\bar{2}2\}}!}{k^{\{1\bar{1}\bar{2}2\}}! \bar{k}^{\{1\bar{1}\bar{2}2\}}!}
P_{12}^{k^{\{1\bar{1}\bar{2}2\}}}(-2 \bar{v}_1\cdot \bar{v}_2)^{\bar{k}^{\{1\bar{1}\bar{2}2\}}}\nn\\
&&\qquad\qquad \times P_{01}^{\bar{k}^{\{1\bar{1}\bar{2}2\}}} P_{02}^{\bar{k}^{\{1\bar{1}\bar{2}2\}}}  (-2 \bar{v}_1\cdot P_0)^{k^{\{1\bar{1}\bar{2}2\}}} (-2 \bar{v}_2\cdot P_0)^{k^{\{1\bar{1}\bar{2}2\}}}.\nn
\eea
For the binomial expansion of this factor, we introduced two non-negative $k^{\{1\bar{1}\bar{2}2\}}$ and $\bar{k}^{\{1\bar{1}\bar{2}2\}}$. 
Similarly, we introduced $k^{\{3\bar{3}\bar{4}4\}}$ and $\bar{k}^{\{3\bar{3}\bar{4}4\}}$ for the the expansion of $\beta^{\{3\bar{3}\bar{4}4\}}$.
The other factors which comes from $\{R\}$ or $\{L\}$ combine into a single factor.
After expanding all the factors and collecting various terms, we obtain:
\bea\label{contraction v}
\mathcal{V}({\bf m}_L, {\bf m}_R)&=&\tilde{\sum_{r,\beta, k}}~~
\prod_{(ij)} \frac{P_{ij}^{\sum_{\{A\}} k_{ij}^{\{A\}}}}{\prod_{\{A\}}k_{ij}^{\{A\}}! }
\prod_{(\bar{i}j )} \frac{(-2\bar{v}_i\cdot P_j)^{\sum_{\{A\}} k_{\bar{i}j}^{\{A\}} }}{\prod_{\{A\}} k^{\{A\}}_{\bar{i}j}! }
\prod_{(\bar{i}\bar{j}) } \frac{(-2\bar{v}_i\cdot \bar{v}_j)^{\sum_{\{A\}} k_{\bar{i}\bar{j}}^{\{A\}} }}{\prod_{\{A\}} k^{\{A\}}_{\bar{i}\bar{j}}! }
\\[5pt]
&&\times
\prod_{i} P_{0i}^{ \kappa_i}(-2\bar{v}_i\cdot P_0)^{\bar{\kappa}_{\bar{i}} }
P_{12}^{r-\bar{k}^{\{1\bar{1}\bar{2}2\}}} P_{34}^{r-\bar{k}^{\{3\bar{3}\bar{4}4\}}}
 (-2\bar{v}_1\cdot \bar{v}_2)^{\bar{k}^{\{1\bar{1}\bar{2}2\}}} (-2\bar{v}_3\cdot \bar{v}_4)^{\bar{k}^{\{3\bar{3}\bar{4}4\}}}
 \nn
\eea
Here $\kappa_i$ and $\bar{\kappa}_{\bar{i}}$ are defined as:
{
\bea\label{def:kappa i}
\kappa_i&=&\sum_{\{A\} (\ni i) } (\beta^{\{A\}}-\sum_{\alpha\in \{A\}}k^{\{A\}}_{i\alpha} )+\sum_{\{B\} (\ni (ii))}\beta^{\{B\}} +\sum_{\{B\}(\ni i)}\bar{k}^{\{B\}}\,,\nn\\
\bar{\kappa}_{\bar{i}}&=&\sum_{ \{A\}(\ni \bar{i})} (\beta^{\{A\}}-\sum_{\alpha\in \{A\}}k_{\bar{i}\alpha}^{\{A\}} ) +\sum_{\{B\}(\ni (ii\bar{i}))}\beta^{\{B\}} +\sum_{\{B\}(\ni i)}k^{\{B\}} \,.
\eea
}
Note that $\{B\}=\{L\} \cup \{R\}$\,, and $\alpha$ runs over $i$ and $\bar{i}$.
The summation $\tilde{\sum}$ is given by:
\bea \label{r beta k sum}
\tilde{\sum_{r,\beta, k}}&=&\tilde{\sum_r}\sum_{\beta^{\{A\}}, \beta^{\{L\}}, \beta^{\{R\}}}\frac{m^+_0!n_{01}!n_{02}!m^-_0! n_{03}!n_{04}!}{(J!)^2}\frac{(J-2r)!}{2^{J-2r}}
 \frac{r!}{\prod_{\{L\}}\beta^{\{L\}}!}\frac{r!}{\prod_{\{R\}}\beta^{\{R\}}!}
 \\
&\times&
\sum_{\sum_{(\alpha \beta)} k_{(\alpha\beta)}^{\{A\}} =\beta^{\{A\}} }
{
\sum_{k^{\{1\bar{1}\bar{2}2\}}, k^{\{3\bar{3}\bar{4}4\}}}
(-1)^{\sum_{\{A\}} k_{(1)(3)}^{\{A\}}+ k_{(2)(4)}^{\{A\}} }
\frac{\beta^{\{1\bar{1}\bar{2}2\}}!}{k^{\{1\bar{1}\bar{2}2\}}!\bar{k}^{\{1\bar{1}\bar{2}2\}}!}\frac{\beta^{\{3\bar{3}\bar{4}4\}}!}{k^{\{3\bar{3}\bar{4}4\}}!\bar{k}^{\{3\bar{3}\bar{4}4\}}!}\,.
}
\nn
\eea
By substituting \eqref{contraction v} into \eqref{general-box} and using the generalized Symanzik star-formula, 
the integration \eqref{general-box} can be expressed into the Mellin representation as:
\bea\label{general integral}
\cI^{({\bf n}_L,{\bf n}_R)}_{\nu, J}(P_i, Z_i)
&=&\pi^h \left(\prod_{i=1}^4\frac{m_i!}{\Gamma (\tilde{\gamma}_{0i})}\right)
 \sum_{\{a,b\}} \int_{-i\infty}^{i\infty} \frac{d\tilde{\delta}_{12}d\tilde{\delta}_{13}}{(2 \pi i)^2}\, \tilde{P}^{(0)(\bn_L,\bn_R)}_{\nu, J}(\tilde{\delta}_{ij},a_{ij},b_{ij})\nn\\
&\times&
\frac{1}{\Gamma(\tilde{\delta}_{12})\Gamma(\tilde{\delta}_{34})}
\prod_{i\neq j} \frac{{(-2\bar{v}_i\cdot P_j)^{a_{ij}}}}{a_{ij}!}
\prod_{i<j}\frac{\left(-2\bar{v}_i\cdot \bar{v}_j\right)^{b_{ij}}}{b_{ij}!} \Gamma(\tilde{\delta}_{ij})P_{ij}^{-\tilde{\delta}_{ij}}
\eea
Now the Mellin variables $\{\tilde{\delta}_{ij}, a_{ij}, b_{ij}\}$ satisfy the conditions;
\bea\label{shifted condition}
{\sum_{j(\neq i)}(b_{ij}+a_{ij})=l_i\,,\quad \sum_{j(\neq i)}(\tilde{\delta}_{ij}-a_{ji})=\Delta_i\,,}
\eea
where $a_{ij} \neq a_{ji}$ and $b_{ij} = b_{ji}$ are non-negative integers.
The range of summation of $a_{ij}$ and $b_{ij}$ restricted by the first constraint,
and the second constraint of $\tilde{\delta}_{ij}$ is shifted from the scalar case \eqref{deltaij-constraint} by $a_{ij}$ and $b_{ij}$.
Here $\tilde{P}^{(0) (\bn_L, \bn_R)}_{\nu;J}$ is
\bea
&& \tilde{P}_{\nu,J}^{(0)({\bf n}_L,{\bf n}_R)}(\tilde{\delta}_{ij},a_{ij},b_{ij})\equiv \tilde{\sum_{r,\beta, k}}\frac{1}{(-2)^{\sum_i m_i +n_{12}+n_{34}}}  \prod_i (\tilde{\gamma}_{0i}-\kappa_i)_{\kappa_i}
(m_i+1)_{\bar{\kappa}_{\bar{i}}}
\nn\\
&&\quad \times
\Gamma(\bar{\delta}_{12})\Gamma(\bar{\delta}_{34})
\frac{b_{12}!}{(b_{12}-n_{12}-\bar{k}^{\{1\bar{1}\bar{2}2\}})!}
\frac{b_{34}!}{(b_{34}-n_{34}-\bar{k}^{\{3\bar{3}\bar{4}4\}})!}
 \nn\\
&&\quad \times 
\prod_{(ij)} \frac{(\tilde{\delta}_{ij})_{\sum_{\{A\}} k_{ij}^{\{A\}}} }{\prod_{\{A\}} k_{ij}^{\{A\}}!} 
\prod_{(\bar{i}j)} \frac{(a_{ij}-\sum_{\{A\}}k_{\bar{i}j}^{\{A\}} +1)_{\sum_{\{A\}} k_{\bar{i}j}^{\{A\}}} }{\prod_{\{A\}} k_{\bar{i}j}^{\{A\}}!} 
\prod_{(\bar{i}\bar{j})}\frac{(b_{ij}-\sum_{\{A\}}k_{\bar{i}\bar{j}}^{\{A\}} +1)_{\sum_{\{A\}} k_{\bar{i}\bar{j}}^{\{A\}}} }{\prod_{\{A\}} k_{\bar{i}\bar{j}}^{\{A\}}!} \,,
\nn\\
\eea
Notice that $\tilde{\delta}_{ij}$ depend on the integers  $a_{ij}$ and $b_{ij}$ due to the condition \eqref{shifted condition}.
We have also defined $\bar{\delta}_{12}$ and $\bar{\delta}_{34}$ as:
\bea 
\bar{\delta}_{12} &\equiv& \tilde{\delta}_{12}-\tilde{\gamma}_{12}+n_{12}+r-\bar{k}^{\{1\bar{1}\bar{2}2\}}\,,\\
\bar{\delta}_{34} &\equiv& \tilde{\delta}_{34}-\tilde{\gamma}_{34}+n_{34}+r-\bar{k}^{\{3\bar{3}\bar{4}4\}}\,.
\eea
In the Mellin representation, the combinations of $P_i$ and $\bar{v}_i$ give the elements of tensor structures 
\bea\label{list P dot v}
\bar{v}_1\cdot P_3 &=&\frac{P_{23}}{P_{12}} V_{1,23}\,,\quad \bar{v}_1\cdot P_4= \frac{P_{24}}{P_{12}}V_{1,24}\,,
\quad \bar{v}_2 \cdot P_3 = -\frac{P_{13}}{P_{12}} V_{2,13}\,,\quad \bar{v}_2 \cdot P_4 = -\frac{P_{14}}{P_{12}}V_{2,14}\,,\no\\
\bar{v}_3\cdot P_1&=&\frac{P_{14}}{P_{34}} V_{3,41}\,,\quad \bar{v}_3 \cdot P_2= \frac{P_{24}}{P_{34}}V_{3,42}\,,
\quad \bar{v}_4\cdot P_1=-\frac{P_{13}}{P_{34}} V_{4,31}\,,\quad \bar{v}_4\cdot P_2= -\frac{P_{23}}{P_{34}}V_{4,32}\,.\no\\
\eea
These are proportional to $V_{i,jk}$ and the products of $\bar{v}_i$ relate to $H_{ij}$ and $V_{i, jk}$
\bea\label{list v dot v}
\bar{v}_1\cdot \bar{v}_2 &=&\frac{H_{12}}{P_{12}}\,,
\quad \bar{v}_1\cdot \bar{v}_3 = \frac{-P_{12}P_{34}H_{13}+2 P_{14} P_{23} V_{1,23}V_{3,41}}{P_{12}P_{34}P_{13}}\,,\\
\bar{v}_3\cdot \bar{v}_4 &=&\frac{H_{34}}{P_{34}}\,,
\quad \bar{v}_1\cdot \bar{v}_4 = \frac{P_{12}P_{34}H_{14}-2 P_{13} P_{24} V_{1,24}V_{4,31}}{P_{12}P_{34}P_{14}}\,,\nn\\
\bar{v}_2\cdot \bar{v}_3 &=& \frac{P_{12}P_{34}H_{23}-2 P_{13} P_{24} V_{2,13}V_{3,42}}{P_{12}P_{34}P_{23}} \,,
\quad \bar{v}_2\cdot \bar{v}_4 = \frac{-P_{12}P_{34}H_{24}+2 P_{14} P_{23} V_{2,14}V_{4,32}}{P_{12}P_{34}P_{24}}\,.\nn
\eea
Now $P_i\cdot \bar{v}_j$ and $\bar{v}_i \cdot \bar{v}_j$ become the elements of tensor structure for the four point case instead of $V_{i,jk}$ and $H_{ij}$\,.
Next we rescale the tensor structures as:  
\bea\label{Def:v and h}
(-2 \bar{v}_i \cdot \bar{v}_j) \rightarrow \tilde{H}_{ij} \equiv \left(\frac{-2 \bar{v}_i \cdot \bar{v}_j}{P_{ij}}\right)\,, \qquad (-2 \bar{v}_i \cdot P_j) \rightarrow  \tilde{V}_{ij}\equiv  \left(\frac{-2 \bar{v}_i \cdot P_j}{P_{ij}}\right)\,.
\eea
Here we defined new tensor structures $\tilde{H}_{ij}$ and $\tilde{V}_{ij}$\,.
After this rescaling, $\tilde{\delta}_{ij}$ are shifted as 
$\tilde{\delta}_{ij}\rightarrow \delta_{ij}$\,,
and \eqref{general integral} becomes \eqref{general integral2}, 
and the constraint for $\delta_{ij}$ are changed as in \eqref{constraint2}.
Now $\bar{\delta}_{12}$ and $\bar{\delta}_{34}$ are:
{
\bea 
\bar{\delta}_{12} &=&\delta_{12}-\tilde{\gamma}_{12}+n_{12}+r-\bar{k}^{\{1\bar{1}\bar{2}2\}}+b_{12}\nn\\
&=&\delta_{12}-\frac{\tilde{\tau}_1 + \tilde{\tau}_2-\tilde{\tau}_0^+}{2}+d_{12}\,,\\
\bar{\delta}_{34} &=&\delta_{34}-\tilde{\gamma}_{34}+n_{34}+r-\bar{k}^{\{3\bar{3}\bar{4}4\}}+b_{34}\nn\\
&=&\delta_{12}-\frac{\tilde{\tau}_3 + \tilde{\tau}_4-\tilde{\tau}_0^-}{2}+d_{34}\,.
\eea
}
where $d_{12}$ and $d_{34}$ are defined as:
\be \label{def:d}
d_{12}= r+b_{12}-n_{12}-\bar{k}^{\{1\bar{1}\bar{2}2\}},\quad
d_{34}= r+b_{34}-n_{34}-\bar{k}^{\{3\bar{3}\bar{4}4\}}.
\ee
Note that $d_{12}$ and $d_{34}$ are non-negative integers 
because due to the factor in the Mack polynomial $\frac{1}{(b_{12}-n_{12}-\bar{k}^{\{1\bar{1}\bar{2}2\}})!} =\frac{1}{\Gamma(b_{12}-n_{12}-\bar{k}^{\{1\bar{1}\bar{2}2\}} +1)} $, 
if $b_{12}-n_{12}-\bar{k}^{\{1\bar{1}\bar{2}2\}}<0$\,, this factor becomes zero.
Therefore only $b_{12}$ which is greater than $n_{12}+\bar{k}^{\{1\bar{1}\bar{2}2\}}$ can contribute, 
and for the same reason, we can regard $d_{34}$ as a non-negative integer. Now we can decompose the gamma functions 
$\Gamma(\bar{\delta}_{12})$ and $\Gamma(\bar{\delta}_{34})$ as:
\be
\Gamma(\bar{\delta}_{12})= \Gamma(\hat{\delta}_{12}) (\hat{\delta}_{12})_{d_{12}},\quad 
\Gamma(\bar{\delta}_{34})= \Gamma(\hat{\delta}_{34}) (\hat{\delta}_{34})_{d_{34}},
\ee
where $\hat{\delta}_{12}$ and $\hat{\delta}_{34}$ are defined in \eqref{hat delta}. The other $\tilde{\delta}_{ij}$ are also shifted as 
\be\label{shift:tilde delta}
\tilde{\delta}_{ij}\rightarrow \delta_{ij} = \tilde{\delta}_{ij}-d_{ij}\,,\quad 
d_{ij}\equiv a_{ij}+a_{ji}+b_{ij}\,. \qquad (\,ij \in (ij)\,)
\ee
Substituting this decomposition  into \eqref{general integral}, we obtain the Mellin representation \eqref{general integral2}.


\section{A Useful Identity for Proving the Equivalence of Mack Polynomials in  \cite{ConformalRT} and \cite{Gopakumar2016-1}}\label{App:Id-Mack}
\paragraph{}
In \cite{ConformalRT}, the authors use a different notation of the Mack polynomial by the factor $\prod_{i=1}^4 (\alpha_i)_{n_i}$, where
\bea
\alpha_i&\equiv& 1-\gamma_{0i}\,, \quad n_i\equiv J-r-\sum_j k_{ij}\,.
\eea
We can show the equivalence, using Eular's reflection identity $\Gamma(z)\Gamma(1-z)= \pi/\sin \pi z$\,.
The last factors in \eqref{Def:tildeP} are rewritten as the following way: 
\bea
(\gamma_{0i}-n_i)_{n_i}= \frac{\Gamma(\gamma_{0i})}{\Gamma(\gamma_{0i}-n_i)}= \frac{\sin \pi(\gamma_{0i}-n_i)}{\sin \pi(\gamma_{0i})}\frac{\Gamma(1-\gamma_{0i}+n_i)}{\Gamma(1-\gamma_{0i})}=(-1)^{n_i}(\alpha_i)_{n_i}\,.
\eea
Now we obtain the alternative expression for the Mack polynomial, replacing the factors:
\bea\label{Id-Mack}
\prod_{i=1}^4(\gamma_{0i}-n_i)_{n_i}=\prod_{i=1}^4(-1)^{n_i}(\alpha_i)_{n_i}=\prod_{i=1}^4(\alpha_i)_{n_i}\,,
\eea
where we used $(-1)^{\sum_{i=1}^4 n_i}=(-1)^{2J}=1$\,.


\bibliographystyle{sort}

\end{document}